\documentclass[fleqn,usenatbib]{mnras}

\usepackage[T1]{fontenc}

\DeclareRobustCommand{\VAN}[3]{#2}
\let\VANthebibliography\thebibliography
\def\thebibliography{\DeclareRobustCommand{\VAN}[3]{##3}\VANthebibliography}

\usepackage{graphicx}	% Including figure files
\usepackage{amsmath}	% Advanced maths commands
\usepackage{amssymb}	% Extra maths symbols
\usepackage{cleveref}
\usepackage{tabularx}
\usepackage[skip=2pt]{caption}
\usepackage{txfonts}
\usepackage{orcidlink}
\usepackage{array}
\usepackage{relsize}
\usepackage{acro}
\usepackage{multicol}
\usepackage{multirow}
\usepackage{xfrac}
\usepackage{lipsum}

\DeclareAcronym{mqg}{
  short = MQG ,
  long  = massive quiescent galaxy ,
  long-plural-form = massive quiescent galaxies ,
  tag = abbrev
}

\DeclareAcronym{coda}{
  short = CoDaIII ,
  long  = Cosmic Dawn III ,
  tag = abbrev
}

\DeclareAcronym{illtng}{
  short = TNG ,
  long  = Illustris: The Next Generation ,
  tag = abbrev
}

\DeclareAcronym{gqt}{
  short = GQT ,
  long  = Gaussian quantile transformation ,
  tag = abbrev
}

\DeclareAcronym{cdf}{
  short = CDF ,
  long  = cumulative distribution function ,
  tag = abbrev
}

\DeclareAcronym{imf}{
  short = IMF ,
  long  = initial mass function ,
  tag = abbrev
}

\DeclareAcronym{sfh}{
  short = SFH ,
  long  = star formation history ,
  long-plural-form = star formation histories ,
  tag = abbrev
}

\DeclareAcronym{zh}{
  short = ZH ,
  long  = metallicity history ,
  long-plural-form = metallicity histories ,
  tag = abbrev
}

\DeclareAcronym{bgs}{
  short = BGS ,
  long  = Bright Galaxy Survey ,
  tag = abbrev
}

\DeclareAcronym{grispy}{
  short = GriSPy ,
  long  = Grid Search In Python ,
  tag = abbrev
}

\DeclareAcronym{disperse}{
  short = DisPerSE ,
  long  = Discrete Persistent Structure Extractor ,
  tag = abbrev
}

\DeclareAcronym{ghc}{
  short = GHC ,
  long  = galaxy-halo connection ,
  tag = abbrev
}

\DeclareAcronym{shmr}{
  short = SHMR ,
  long  = stellar-halo mass relation ,
  tag = abbrev
}

\DeclareAcronym{mzr}{
  short = MZR ,
  long  = mass-metallicity relation ,
  tag = abbrev
}

\DeclareAcronym{mhzr}{
  short = HMZR ,
  long  = halo mass-metallicity relation ,
  tag = abbrev
}

\DeclareAcronym{ssp}{
  short = SSP ,
  long  = simple stellar population ,
  tag = abbrev
}

\DeclareAcronym{srcc}{
  short = $r_S$ ,
  long  = Spearman's rank correlation coefficient ,
  tag = abbrev
}

\DeclareAcronym{mwa}{
  short = MWA ,
  long  = mass-weighted age ,
  tag = abbrev
}

\DeclareAcronym{zwa}{
  short = ZWA ,
  long  = metallicity-weighted age ,
  tag = abbrev
}

\DeclareAcronym{mpb}{
  short = MPB ,
  long  = main progenitor branch ,
  long-plural-form = main progenitor branches ,
  tag = abbrev
}

\DeclareAcronym{rfr}{
  short = RFR ,
  long  = random forest regressor ,
  tag = abbrev
}

\DeclareAcronym{ert}{
  short = ERT ,
  long  = extremely randomised tree ,
  tag = abbrev
}

\DeclareAcronym{fof}{
  short = FoF ,
  long  = Friends-of-Friends ,
  tag = abbrev
}

\DeclareAcronym{relu}{
  short = ReLU ,
  long  = rectified linear unit ,
  tag = abbrev
}

\DeclareAcronym{lrelu}{
  short = L-ReLU ,
  long  = leaky rectified linear unit ,
  tag = abbrev
}

\DeclareAcronym{elu}{
  short = ELU ,
  long  = exponential linear unit ,
  tag = abbrev
}

\DeclareAcronym{fsps}{
  short = FSPS ,
  long  = Flexible Stellar Population Synthesis ,
  tag = abbrev
}

\DeclareAcronym{UM}{
  short = UM ,
  long  = UniverseMachine ,
  tag = abbrev
}

\DeclareAcronym{desi}{
  short = DESI ,
  long  = Dark Energy Spectroscopic Instrument ,
  tag = abbrev
}

\defcitealias{Chittenden}{CT23}
\defcitealias{Behera}{BTC25}
\title[Galaxy Formation In Dark Matter Simulations]{Evaluating the galaxy formation histories predicted by a neural network in pure dark matter simulations}

\author[Chittenden, Behera \& Tojeiro 2025]{
Harry George Chittenden,$^{\orcidlink{0000-0001-5856-8713} \ 1,2}$\thanks{E-mail: hchittenden@swin.edu.au}
Jayashree Behera$^{\orcidlink{0009-0002-2434-5903} \ 3}$
and Rita Tojeiro$^{\orcidlink{0000-0001-5191-2286} \ 2}$ 
\\
$^{1}$Centre for Astrophysics and Supercomputing [CAS], Swinburne University of Technology, P.O. Box 218, Hawthorn VIC 3122, Melbourne, Australia\\
$^{2}$School of Physics \& Astronomy, University of St Andrews, North Haugh, St Andrews KY16 9SS, Scotland, United Kingdom\\
$^{3}$Department of Physics, Kansas State University, Cardwell Hall, 116, 1228 N M.L.K. Jr. Dr, Manhattan, KS 66506, United States
}

\date{Accepted XXX. Received YYY; in original form ZZZ}

\pubyear{2025}

\begin{document}
\label{firstpage}
\pagerange{\pageref{firstpage}--\pageref{lastpage}}
\maketitle

\begin{abstract}
We investigate a series of galaxy properties computed using the merger trees and environmental histories from dark matter only cosmological simulations, using a semi-recurrent neural network producing self-consistent predictions of galaxy evolution, and using stochastic improvements to this model based on similarly predicted Fourier Transforms. We apply these methods to the dark matter only runs of the IllustrisTNG simulations to understand the effects of baryon removal, and to the gigaparsec-volume pure dark matter simulation Uchuu, to understand the effects of the lower resolution or alternative metrics for halo properties. We find that the machine learning model recovers accurate summary statistics derived from the predicted star formation and stellar metallicity histories, and correspondent spectroscopy and photometry. However, the inaccuracies of the model’s application to dark simulations are substantial for low mass and slowly growing haloes. For these objects, the halo mass accretion rate is exaggerated due to the lack of stellar feedback, yet the formation of the halo can be severely limited by the absence of low mass progenitors in a low resolution simulation. Furthermore, differences in the structure and environment of higher mass haloes results in an overabundance of red, quenched galaxies. These results signify progress towards a machine learning model which builds high fidelity mocks based on a physical interpretation of the galaxy-halo connection, yet they illustrate the need to account for differences in halo properties and the resolution of the simulation.
\end{abstract}


\section{Introduction}
\label{sec:intro}

Cosmic hydrodynamical simulations are a valuable tool for modelling the causal relationship between galaxies and haloes across cosmic time; known as the \ac{ghc}. Modelling galaxy evolution in its full complexity, however, is a difficult task. Numerous physical processes such as the accretion and cooling of gas are necessary to fuel star formation and synthesise heavy elements \citep{Somerville, Vogelsberger}, which in turn are influenced by the growth of the dark matter halo and its interactions with the surrounding cosmological structures \citep{Wechsler}. Capturing the complete \ac{ghc} requires details of large scale structure evolution on megaparsec scales to compute massive galaxies and clusters, and the physics internal to the galaxy to compute summary statistics such as mass-metallicity relations.

In \citet{Chittenden}, hereafter \citetalias{Chittenden}, we designed a semi-recurrent neural network capable of reproducing the \ac{ghc} in the \ac{illtng} simulations \citep{Nelson2018, IllustrisTNG, Pillepich2017, Springel, Marinacci, Naiman}, in which we predict the \ac{sfh} and \ac{zh} of central and satellite galaxies, from which we recover physical relations such as the \ac{shmr} and \ac{mzr}, reproduce observational statistics such as colour bimodality, and show the growth and internal properties of the halo to influence the predicted \ac{sfh} and stellar mass, while environmental properties influence the \ac{zh} and stellar metallicity.

Machine learning models which can emulate galaxy properties from their dark matter component alone can be used to make predictions in cosmic N-body simulations, where the absence of galaxies allows for such simulations to computed with much greater size and resolution than their hydrodynamical counterparts. Future studies may benefit from a simulation containing galaxy evolution on the scales captured by a high volume, high resolution simulation; however, the limited computational resources required for such simulations make computing such a hydrodynamical suite highly impractical.

In this paper, we explicitly evaluate the performance of the trained galaxy evolution model on N-body simulations. Unlike previous studies which compute galaxy statistics in N-body simulations using common methodology such as U-Net convolutional networks \citep{Chadayammuri, Caro, DarkAI}, the \citetalias{Chittenden} model employs a semi-recurrent neural network design; incorporating recurrent connections which allow the model to retain memory of the halo and environmental properties across sequential time steps, which has shown to significantly advance the training speed and predictive accuracy of the network.

While studies such as \citet{Xu2,McGibbonKhochfar,McGibbonKhochfar2} incorporate selected historical information about the dark matter halo, this model encodes the full formation histories of each halo while strictly enforcing causal predictions of the state of its galaxy, thereby making self-consistent predictions of star formation and metallicity histories based on the complete history of the halo, while quantities such as stellar mass, metallicity and photometric colours are calculated from these star formation and metallicity histories; a feature not utilised in most machine learning models trained on cosmic simulations. We also derive observable quantities and summary statistics by forward modelling our predictions, making all results of this study self-consistent, which \citetalias{Chittenden} emphasise as necessary for the \ac{ghc} to be inferred from mock observations. This paper marks the first implementation of a semi-recurrent model to an N-body simulation on which it was not previously trained.

\begin{table}
\begin{center}
\begin{tabular}{|c|c|c|c|c|}
\hline
 \multicolumn{5}{|c|}{Simulation Properties} \\ 
 \hline \hline
 Suite & Simulation & $L_\text{box}$ (Mpc) & $m_\text{DM}$ ($M_\odot$) & $N_\text{snap}$ \\ \hline \hline
 TNG - & TNG100-1 & $110.7174$ & $7.5 \times 10^6$ & $100$ \\ \cline{2-5}
 Hydro & TNG300-1 & $302.6277$ & $5.9 \times 10^7$ & $100$ \\ \hline \hline
 TNG - & TNG100-1-Dark & $110.7174$ & $8.9 \times 10^6$ & $100$ \\ \cline{2-5}
 Dark & TNG300-1-Dark & $302.6277$ & $7.0 \times 10^7$ & $100$ \\ \hline \hline
 Uchuu & Uchuu & $2952.4653$ & $4.8 \times 10^8$ & $50$ \\ \hline
\end{tabular}
\end{center}
\caption{Properties of the simulations used in this work, relating to their volume and resolution. $L_\text{box}$ indicates the side length of the comoving cubic volume of the simulation, $m_\text{DM}$ represents the mass of a single dark matter particle, i.e. the smallest possible mass in the simulation, and $N_\text{snap}$ represents the total number of time snapshots in the simulation. Simulations are grouped in this table according to the suite of simulations (e.g. TNG-Dark) from which they originate. For all of these simulations, mass resolution is decreased as the volume of the simulation and number of particles are increased.}
\label{tab:sims}
\end{table}

Predictions are expected to be somewhat different in dark simulations due to the lack of baryonic effects on the halo mass function \citep{Castro, Anbajagane, Haggar, Riggs}, alternative definitions or calculations of key halo and environmental properties, and differences in mass and spacial resolution; which, in \cref{tab:sims}, we show for the simulations used in this work. We compare predictions in the hydrodynamical \ac{illtng} simulations, discussed at length in \citetalias{Chittenden}, with the equivalent dark simulations in the \ac{illtng} suite (hereafter TNG-Dark), and the Uchuu dark matter simulation \citep{Uchuu}.

Uchuu is a high fidelity N-body simulation which assumes the same \citet{Planck} cosmological parameters as all \ac{illtng} simulations, therefore there are little to no effects to consider regarding mean matter density or Hubble expansion. By comparing predictions in TNG-Hydro with TNG-Dark, we isolate the effect that baryons have on the predicting power of the neural network. By comparing TNG-Dark with Uchuu, we show the effect of Uchuu's lower resolution and alternative definitions of halo properties. The predictions made in Uchuu are therefore the simplest results to be expected for a high fidelity galaxy catalogue produced using our model.

Further to the application of our semi-recurrent predictive model to unseen simulation data, we apply a stochastic modification to the N-body simulation predictions, whose methods and improvements to our TNG-Hydro-based galaxy-halo statistics are outlined in \citet*{Behera}, hereafter \citetalias{Behera}. It is shown in \citetalias{Chittenden} that the machine learning model alone is inadequate for predicting star formation events on short timescales, in spite of the success of predicting the general evolution of galaxies across all times. These "stochastic" features are well recovered by this correction, and the results of \citetalias{Behera} illustrate that certain missing properties of our fiducial galaxy-halo statistics can be attributed to the physics of variable star formation. We apply this modification to the TNG-Dark and Uchuu data to assess the plausibility of reproducing this variability in N-body mocks.

In this paper, we investigate the quality of predictions between our original results and the predictions in TNG-Dark and Uchuu, explain systematic differences between the results in each of these simulation suites and discuss the suitability of our model for reproducing galaxies in a pure dark matter simulation. In \Cref{sec:acq} we discuss the definitions of the dark matter data in each of the simulations, and our methods of acquiring the necessary input data. We compare properties of the input data in \Cref{sec:halos}, baryonic predictions in \Cref{sec:pred} and observational results in \Cref{sec:obs}, eludicating their disparities each time, and exhibiting the improvements made by the stochastic amendment introduced in \citetalias{Behera}. We evaluate the successes and failures of the model, the stochastic correction, and their proficiency for modelling the \ac{ghc} on gigaparsec scales in \Cref{sec:disc}, and summarise our findings in \Cref{sec:conc}.

\section{Data Acquisition}
\label{sec:acq}

In this section, we describe how the data from the pure dark matter simulations are processed for use in our neural network, and discuss the effects of fundamental differences between these simulations on the quality of our data.

Full details regarding data processing are given in \citetalias{Chittenden}, however the data processing procedure can be summarised as follows:

\begin{itemize}
\item Calculate mass accretion histories by finite differencing the (sub)halo mass along the main progenitor branch (MPB).
\item Interpolate the MPB over the time domain of every third snapshot in \ac{illtng}, down to and including $z=0$.
\item For each of these snapshots, calculate overdensities and skews\footnote{A measure of the mass-weighted radial distribution of subhaloes exterior to the target subhalo, tracing subhalo interaction history, as discussed in \citetalias{Chittenden}.} using the \ac{grispy} package \citep{Grispy}.
\item Compute the proxy for circular velocity history: the square root of the ratio of (sub)halo mass ($M_\odot$) to half-mass radius (Mpc).
\item For satellites, compute infall parameters, e.g. scaled infall time, infall velocity, based on and including the scaled accretion time from \citet{Shi}; from the MPBs of the satellite and its host halo. If the simulation contains baryonic data, discard any samples where the stellar mass of the host halo is greater than $10\%$ of the halo mass.
\item For centrals, acquire the $z=0$ cosmic web properties calculated using the \ac{disperse} code \citep{Disperse}. The TNG-Hydro results are publicly available\footnote{\href{https://github.com/Chris-Duckworth/disperse_TNG}{https://github.com/Chris-Duckworth/disperse\textunderscore TNG/}}.
\item Calculate the specific mass accretion gradients shown in \citet{Montero-Dorta}. Fit a Gaussian function to the distribution of gradient values, and discard any samples whose gradient is over $5\sigma$ from the mean of the best fit Gaussian. This quality cut is described further in \cref{sec:halos}.
\item Transform the data to the normalised parameter space seen in \citetalias{Chittenden}, by interpolation between the TNG-Hydro data and the normalised data. Apply this transformed data to the network. Where necessary, a \ac{gqt} was used to normalise the TNG-Hydro data, using time-independent transformations for most temporal variables. The interpolation method made it straightforward to transform new data according to the original mapping between physical and numerical quantities.
\end{itemize}

This gives a list of quantities used in the neural network model, summarised in \cref{tab:networks}. As in \citetalias{Chittenden}, we use the integrated stellar masses and mass-weighted metalllicities derived from predicted \ac{sfh}s and \ac{zh}s to produce summary statistics such as the \ac{shmr} and \ac{mzr}, and use the \ac{fsps} code \citep{ConroyGunnWhite, ConroyGunn} to calculate SEDs and other observables from these network output quantities.

\begin{table}
\centering
\begin{tabular}{|c| m{0.16\textwidth} |m{0.05\textwidth}|m{0.08\textwidth} |c|c|c|c|c|} 
 \cline{2-5}
 \multicolumn{1}{c|}{} & \multicolumn{4}{c|}{Network Data} \\ 
 \cline{2-5}
 \multicolumn{1}{c|}{} & \centering Quantity & \centering Notation & \centering Units & Network  \\ 
 \cline{2-5} \hline
 \parbox[t]{2mm}{\multirow{15}{*}{\rotatebox[origin=c]{90}{Temporal Features}}}
 & \centering \mbox{Halo Mass} \mbox{Accretion Rate} & \centering $\dot{M_h}$ & \centering $M_\odot/\text{Gyr}$ & Both  \\ \cline{2-5}
 & \centering \mbox{Subhalo Mass} \mbox{Accretion Rate} & \centering $\dot{m_h}$ & \centering $M_\odot/\text{Gyr}$ & Satellite  \\ \cline{2-5}
 & \centering 1Mpc \mbox{Overdensity} & \centering $\delta_1$ & & Both  \\ \cline{2-5}
 & \centering 3Mpc \mbox{Overdensity} & \centering $\delta_3$ & & Central  \\ \cline{2-5}
 & \centering 5Mpc \mbox{Overdensity} & \centering $\delta_5$ & & Central  \\ \cline{2-5}
 & \centering \mbox{Circular Velocity} (proxy) & \centering $\tilde{v}_\text{vir}$ & \centering $\sqrt (M_\odot/\text{Mpc})$ & Both  \\ \cline{2-5}
 & \centering \mbox{Dark Matter} \mbox{Half-Mass Radius} & \centering $R_\frac{1}{2}$ & \centering Mpc & Both  \\ \cline{2-5}
 & \centering 1Mpc \mbox{Radial Skew} & \centering $\mu_3$ & & Satellite  \\ \cline{2-5}
 & \centering 3Mpc \mbox{Radial Skew} & \centering $\mu_3$ & & Central  \\ \cline{2-5}
 & \centering \mbox{Distance To} \mbox{Closest Subhalo} & \centering $d_{\mu_3}$ & \centering Mpc & Both  \\ \cline{2-5} \hline \hline
 \parbox[t]{2mm}{\multirow{22}{*}{\rotatebox[origin=c]{90}{Non-Temporal Features}}}
 & \centering \mbox{Specific Halo} \mbox{Mass Accretion} Gradient & \centering $\beta$ (c) \newline $\beta_\text{halo}$ (s) & \centering $\log \text{Gyr}^{-2}$ & Both  \\ \cline{2-5}
 & \centering \mbox{Specific Subhalo} \mbox{Mass Accretion} Gradient & \centering $\beta_\text{sub}$ & \centering $\log \text{Gyr}^{-2}$ & Satellite  \\ \cline{2-5}
 & \centering Scaled Infall Time & \centering $a_\text{infall}$ & & Satellite  \\ \cline{2-5}
 & \centering \mbox{Scaled Formation} Time & \centering $a_\text{max}$ & & Satellite  \\ \cline{2-5}
 & \centering Infall Mass Ratio & \centering $\mu$ & & Satellite  \\ \cline{2-5}
 & \centering Infall Velocity & \centering $v_\text{rel}$ & \centering km/s & Satellite  \\ \cline{2-5}
 & \centering $z=0$ Cosmic Web Distances & \centering $d_\text{CW}$ & \centering kpc & Central  \\ \cline{2-5}
 & \centering Starting Time & \centering $t_\text{start}$ & \centering Gyr & Both  \\ \cline{2-5}
 & \centering \mbox{$z=0$ Halo Mass} & \centering $M_h$ & \centering $M_\odot$ & Both  \\ \cline{2-5}
 & \centering Maximum \mbox{Absolute Halo} \mbox{Accretion Rate} & \centering $\mid\dot{M_h}\mid$ & \centering $M_\odot/\text{Gyr}$ & Both  \\ \cline{2-5}
 & \centering \mbox{$z=0$ Subhalo Mass} & \centering $m_h$ & \centering $M_\odot$ & Satellite  \\ \cline{2-5}
 & \centering Maximum \mbox{Absolute Subhalo} \mbox{Accretion Rate} & \centering $\mid\dot{m_h}\mid$ & \centering $M_\odot/\text{Gyr}$ & Satellite  \\ \cline{2-5} \hline \hline
 \parbox[t]{2mm}{\multirow{7}{*}{\rotatebox[origin=c]{90}{Targets}}}
 & \centering \mbox{Star Formation} History & \centering $\mathcal{S}$ & \centering $M_\odot/\text{Gyr}$ & Both  \\ \cline{2-5}
 & \centering \mbox{Metallicity} \mbox{History} & \centering $\mathcal{Z}$ & \centering $Z_\odot$ & Both  \\ \cline{2-5}
 & \centering \mbox{$z=0$ Stellar} \mbox{Metallicity} & \centering $Z$ & \centering $Z_\odot$ & Both   \\ \cline{2-5}
 & \centering $z=0$ Stellar Mass & \centering $M_s$ & \centering $M_\odot$ & Both   \\ \cline{2-5}
 & \centering \mbox{Mass Weighted Age} & \centering MWA & \centering Gyr & Both  \\
 \hline
\end{tabular}
\caption{Simplification of Table 1 in CT23, summarising the quantities used in both neural networks, according to their input and output layers and their arrangement in the network. Additional columns show the variables' units, and the central and/or satellite networks which use them.}
\label{tab:networks}
\end{table}

\subsection{TNG-Dark}
\label{sec:acqtngdark}

\Cref{tab:sims} shows the key differences between the simulations used in this work. The TNG-Dark simulations are of similar resolution to their hydrodynamical counterparts, and therefore resolution effects will be small if at all applicable. The time domains of their snapshots are also equivalent. The crucial difference between the two results is therefore due to the absence of baryonic effects on the haloes in the TNG-Dark simulations.

In order to interpret the differences between the dark and hydrodynamical \ac{illtng} simulations, we select samples in the dark simulation by cross-matching SubLink trees with our original, hydrodynamical dataset \citep{Nelson2015, Rodriguez-Gomez}. Samples with no cross-matched result in TNG-Dark are discarded. Generally, objects in one simulation are not accurately matched to the other for two reasons. Firstly, objects in proximity to much larger haloes may pass within the virial radius of the larger halo in one of the two simulations, allowing for central haloes to be matched to satellites, and vice versa. Similarly, and particularly so for low mass objects, the SubFind algorithm may combine two distinct subhaloes into a single object, such that one low mass subhalo will cease to exist in the lower resolution simulation (TNG-Dark). For our sample with a minimum stellar mass of $10^{9} M_\odot$ as in \citetalias{Chittenden}, these effects eliminate $\sim 39\%$ of our satellite subhaloes compared with $\sim 1\%$ of our central haloes.

Most TNG-Dark variables were computed in the same manner as the data from hydrodynamical simulations. The exception is the $z=0$ \ac{disperse} cosmic web distances, which, due to the geometric correspondence of the hydrodynamical and dark simulations, are identical to their cross-matched counterparts. We therefore assign the cosmic web properties of TNG-Hydro samples to their cross-matched equivalents in TNG-Dark, in lieu of publicly accessible data for the latter.

\subsection{Uchuu}
\label{sec:acquchuu}

\subsubsection{Merger Forests}
\label{sec:uchuuforests}

Uchuu merger trees are grouped into 2,000 "Forests": ensembles of merger trees which contain all haloes which have interacted with any given member of the forest, and occupy a volume of space separate from all other forests. Each forest can therefore be treated independently of another, as the Consistent-Trees code \citep{ConsistentTrees} used to define merger trees in Uchuu is run independently in groups which occupy a fixed volume, which are then concatenated if they interact or pass within 25 Mpc / $h$ of each other \citep{Uchuu}.

\begin{figure}
\includegraphics[width=\linewidth]{"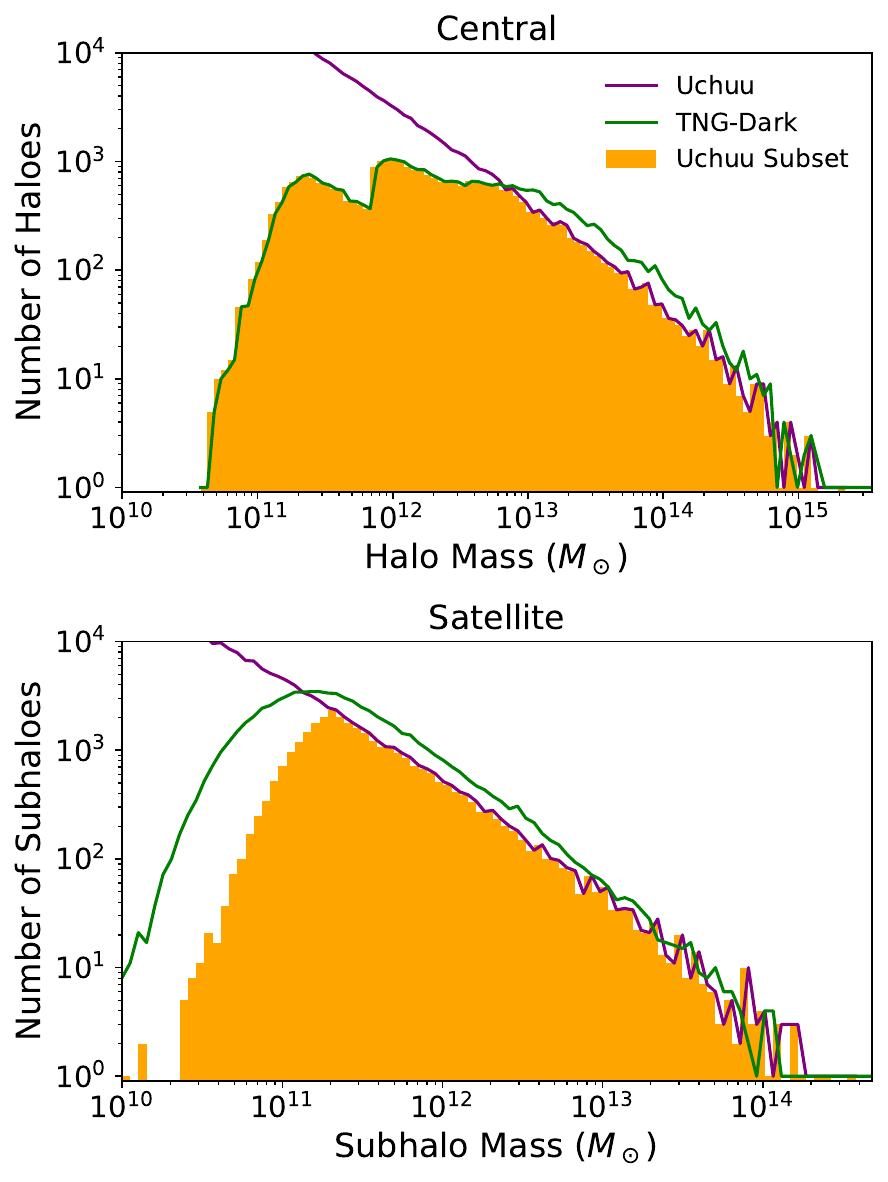"}
\vspace*{-20pt}
\caption{Distributions of central halo and satellite subhalo masses in the complete Uchuu forest (purple) and our cross-matched TNG-Dark sample (green). Sampling the former distribution according to the latter, while imposing quality cuts which invalidate most low-mass Uchuu haloes, we obtain the distribution of Uchuu haloes used in this work (orange). The distribution of central haloes is similar to the TNG-Dark data, however the lack of well-resolved satellite subhaloes at low mass serves to bias the sample distribution of satellite subhaloes in our Uchuu dataset. The 'bump' in the TNG-Dark curve corresponds to the transition between TNG100-1 and TNG300-1 galaxies, given the minimum halo mass of $10^{11.5} M_\odot$ in the latter dataset.}
\label{fig:histmatch}
\end{figure}

In this work we use forest 1411, the largest Uchuu forest, and acquire similar samples to the \ac{illtng} dataset by drawing from the TNG-Dark halo mass distribution and sampling the Uchuu forest at $z=0$ accordingly, provided that they too meet our quality selection criteria. \Cref{fig:histmatch} shows the final result, where it can be shown that the Uchuu forest provides enough well-resolved halo formation histories to match our TNG sample, but does not provide the same quantity of low mass satellite haloes, whose abundance declines sharply below $\sim 4 \times 10^{10} M_\odot$. Following all of our selection procedures, the Uchuu dataset contains approximately 30,000 central and satellite haloes each.

As the chosen Uchuu forest has a geometric mean side length of approximately 234 Mpc, it is of similar volume to the \ac{illtng} simulations, and thus for high mass objects, sampling the (sub)halo mass distributions effectively approximates the (sub)halo mass functions of Uchuu. At lower masses, sampling Uchuu according to the TNG-Dark distribution ensures that the properties in different mass ranges can be compared without sample size bias. While this selection constitutes our test data for the neural network, environmental quantities are computed using the complete Uchuu forest, and are considered unbiased due to the \ac{illtng} and Uchuu simulations having the same cosmic matter density parameters. Furthermore, the total mass densities of individual Uchuu forests differ from their mean by no more than 0.08 dex, being much smaller than the variance in halo overdensities, suggesting that there is no significant environmental bias from any particular choice of Uchuu forest.

We use the YTree Python package to extract the main progenitor branches from the Uchuu forest. The quantities we obtain directly from the forest include halo mass, half mass radius, positions and peculiar velocities. Cosmic web distances are obtained by applying \ac{disperse} to the forest, which being a self-consistent subset of Uchuu means that the algorithm does not need to be applied to the full Uchuu simulation. All other quantities, such as overdensities, are calculated using the same methods as in \citetalias{Chittenden}.

\subsubsection{Alternative Definitions}
\label{sec:uchuudef}

Unlike \ac{illtng}, SubLink merger trees do not exist in Uchuu, and haloes and their substructures are defined using the Rockstar code \citep{Rockstar}. There exists a flag in each Uchuu merger tree which indicates the ID of the halo which hosts the target halo. A "first order" satellite halo is one whose host is a central "zeroth order" halo, a "second order" halo is hosted by a first order halo, and so on. We therefore treat first order haloes as satellite haloes, and zeroth order haloes as central haloes.

In \ac{illtng}, halo and subhalo masses are taken as the sum of all masses of dark matter particles bound to the group by the SubLink algorithm. This field does not exist in the Uchuu merger trees, however a number of definitions of mass exist. We find that the closest match to this field which exists in both suites is $M_{200\text{c}}$: the total mass enclosed within an overdense region equal to 200 times the critical mass density of the universe. These mass accretion histories show similar behaviour in TNG-Dark and Uchuu, and produce a similar overall mass distribution, and consequently the predictions of the neural network are similar when trained using $M_{200\text{c}}$ and with the SubLink mass. This is our choice of field representing halo masses in Uchuu, given its strong correlation with the total particle masses used in \citetalias{Chittenden}. Halo mass dependent calculations such as overdensities are also calculated using this quantity.

\subsubsection{Resolution Differences}
\label{sec:uchuures}

\Cref{tab:sims} shows that the mass resolution of Uchuu is an approximate order of magnitude weaker than the TNG300 simulations. This results in very poor resolution of low mass haloes in Uchuu, and the absence of low mass, first order satellite haloes, following our quality cuts, serves to limit the number of satellites below a subhalo mass of $\sim 2 \times 10^{11} M_\odot$; which we show in \cref{fig:histmatch}. We are left with a modest sample of low mass satellite haloes, but a clearly different distribution of subhalo masses below $\sim 2 \times 10^{11} M_\odot$.

Another important detail is the difference in time resolution of the Uchuu snapshots compared with \ac{illtng}. Uchuu has half as many snapshots as \ac{illtng}, as shown in \cref{tab:sims}, however the temporal separation of snapshots is smaller than \ac{illtng} for $2\lesssim z\lesssim 4$ and larger otherwise. All temporal properties, regardless of their simulation, are linearly interpolated over the same 33 snapshots in \ac{illtng}, yet the sparse nature of these snapshots means that information on short timescales may be lost.

\section{Halo Properties}
\label{sec:halos}

For each simulation used in this study, we apply the same algorithm to acquire the input data, however there exist key differences between the definitions of these properties, depending on the model in question; and differences in their statistics due to the removal of baryons or the resolution of the data. In this section, we discuss the similarities and differences between key input properties of the neural network in each simulation, and how this may influence the predictions of the neural networks.

We compare properties for central and satellite haloes in bins of different mass and accretion history to investigate these differences in separate evolutionary regimes. For centrals, we characterise halos of different accretion histories according to the specific mass accretion gradient $\beta$, defined as the best fit gradient to the following approximate formula for the halo mass $M_h$ as a function of cosmic time $t$, quoted in gigayears:

\begin{equation}
\log_{10} \left( \frac{\dot{M}_h}{M_h} \right) \simeq \gamma + \beta \log_{10} t
\label{eq:beta}
\end{equation}

where $\gamma$ is the constant intercept value at the vertical axis, i.e. where $t=1$ Gyr. As shown in \citetalias{Chittenden}, these $\beta$ values are Gaussian distributed, and so, defining $\alpha$ as the standardised version of $\beta$:

\begin{equation}
\alpha = \frac{\beta - \mu_\beta}{\sigma_\beta}
\label{eq:alpha}
\end{equation}

where $\mu_\beta$ and $\sigma_\beta$ are the best fit parameters of this distribution, we discard low quality samples from all of our data where $\beta$ is more than $5\sigma_\beta$ from the mean, i.e. where $\left| \alpha \right| > 5$. This was also done in \citetalias{Chittenden}, as poor mass or time resolution of a mass assembly history result in poor, unphysical fits to the accretion gradient, deviating significantly from the Gaussian distribution.

\citet{Montero-Dorta} show $\beta$ to be clearly correlated with gas fraction, quenching timescale and assembly bias in TNG300. We use this quantity for quality cuts in both central and satellite datasets, however it is not so useful for modelling satellite subhalo histories due to the distinct nature of the modes of accretion in their central and satellite phases. For satellites, we use the scaled accretion time $a_\text{max}$, which, like $\beta$, is derived explicitly from the subhalo mass accretion history, and is shown by \citet{Shi} to have similar connections with satellite galaxy properties.

In the following figures which compare haloes of different histories, subplots are arranged such that each column is a quintile of (sub)halo mass, with higher mass bins towards the right; while each row is a quartile of accretion gradient, with the steepest accretion histories on the top row. For satellites, higher $a_\text{max}$ values are placed on the top row, such that in both cases, haloes with the earliest half-mass formation times are placed on the top row. These percentiles are taken from the TNG-Hydro data. This convention is adopted for all figures of this format, such that in each case the earliest-forming, gas-poor halos or subhalos appear on the top row. For simplicity, we show only odd-numbered mass quintiles, and only the first and fourth quartiles of mass accretion gradient.

\subsection{Mass Accretion History}
\label{sec:mhdot}

\begin{figure*}
\includegraphics[width=\linewidth]{"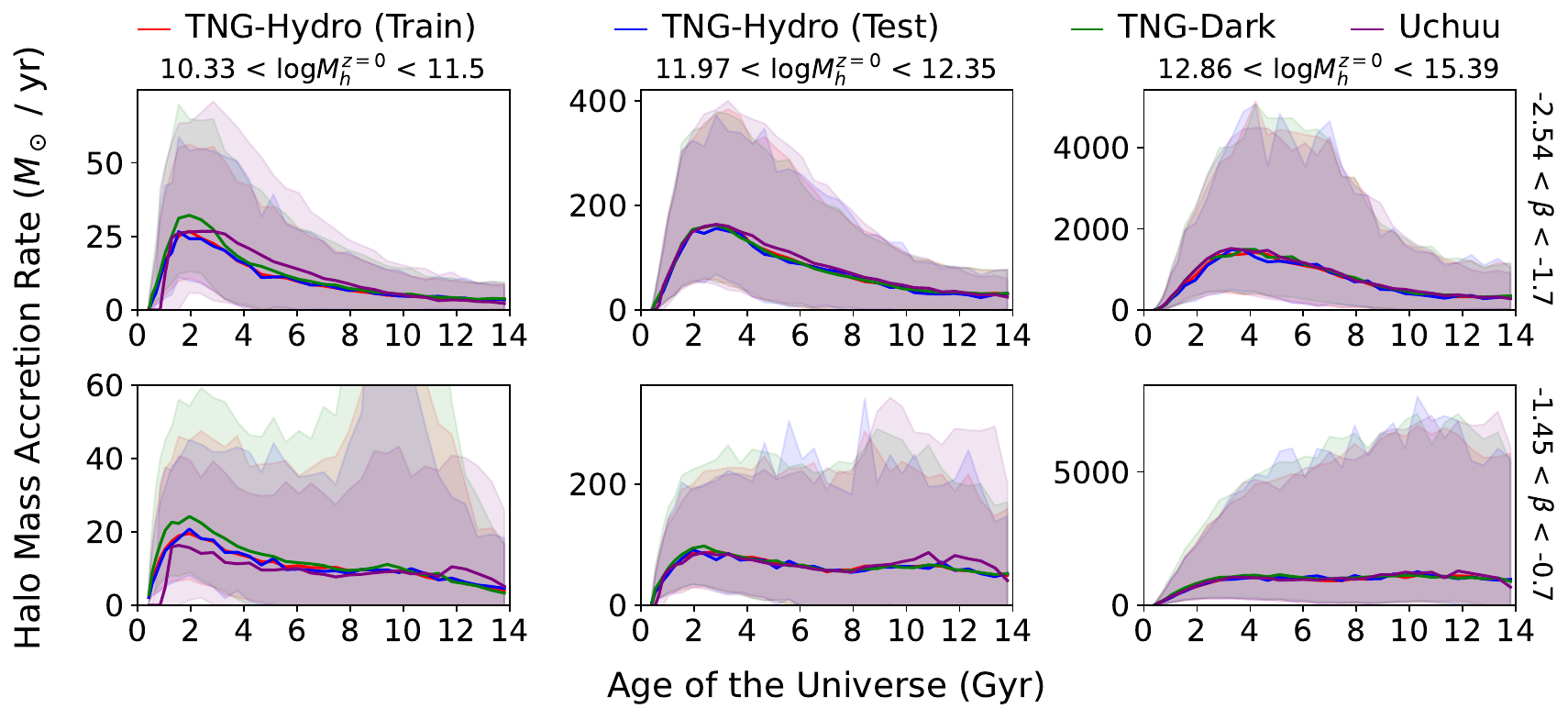"}
\vspace*{-20pt}
\caption{Mass accretion histories of central haloes in bins of halo mass, increasing along the horizontal axis, and specific mass accretion gradient, decreasing in steepness down the vertical axis. Solid lines show the median mass accretion history per bin, while shaded regions show the $15^\text{th}$ and $85^\text{th}$ percentiles. Mass accretion histories are shown for the training (red) and testing (blue) datasets of the TNG-Hydro simulations, alongside TNG-Dark (green) and Uchuu (purple).}
\label{fig:Mhdot}
\end{figure*}

\begin{figure*}
\includegraphics[width=\linewidth]{"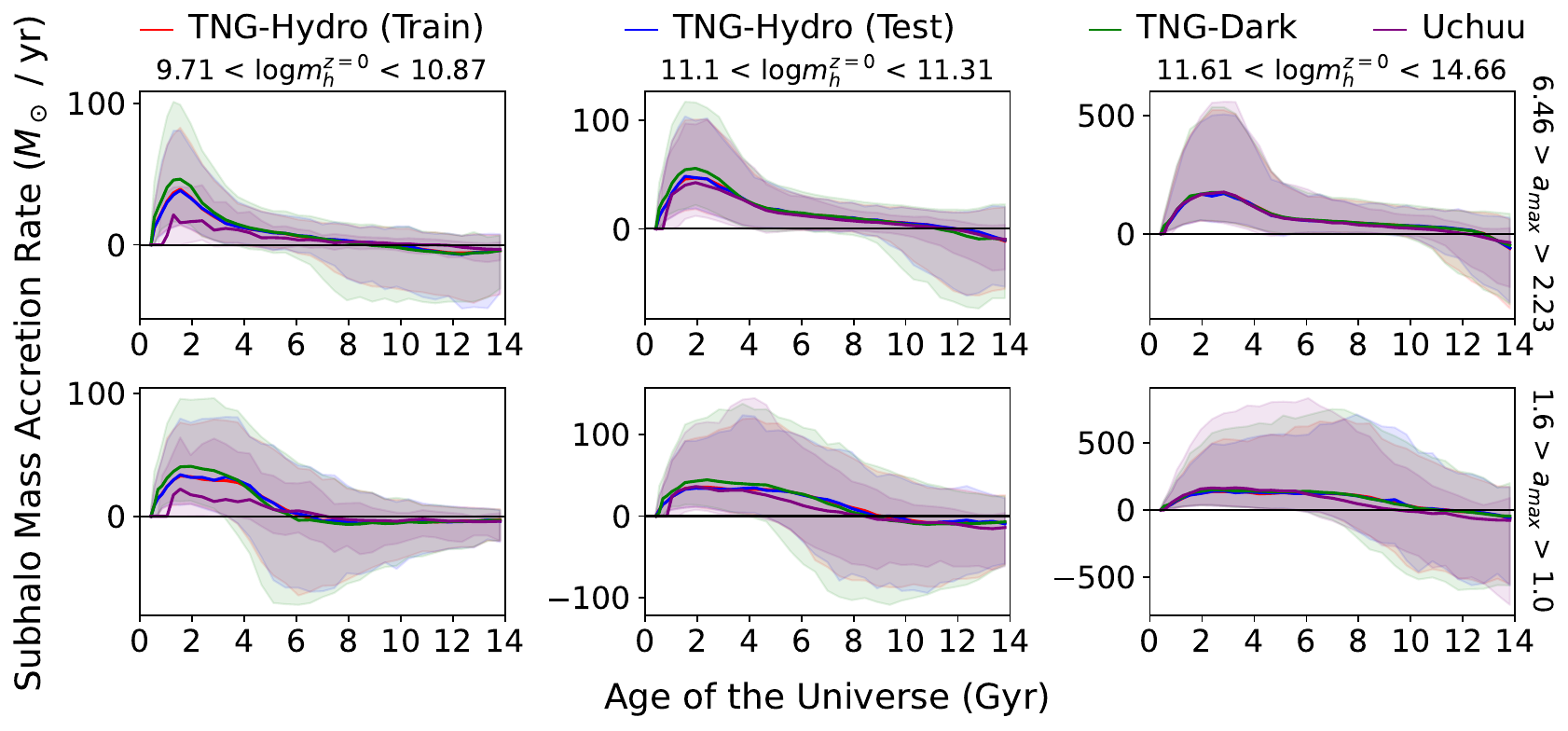"}
\vspace*{-20pt}
\caption{Mass accretion histories of satellite subhaloes, binned in the same manner as in \cref{fig:Mhdot}, using satellite subhalo mass in place of central halo mass, and scaled accretion time in place of specific mass accretion gradient. One key difference in the properties of these accretion histories is their approach towards zero or negative accretion, which rarely happens for central haloes. Subplots show the different times at which subhalo growth stops in different growth regimes.}
\label{fig:mhdot}
\end{figure*}

\Cref{fig:Mhdot,fig:mhdot} show the median and $15^\text{th}-85^\text{th}$ percentile ranges of mass accretion histories in bins of final (sub)halo mass and mass accretion gradient. In most bins, these mass accretion histories are very similar. The differences between simulations arise in low mass and shallow gradient cases, where the amplitude of Uchuu mass accretion histories are reduced, and in the low mass regime is increased for TNG-Dark haloes at early times.

The enhancement of mass for early, low mass haloes is likely due to the abscence of baryonic driven outflows, which will have a substantial effect on haloes of a high gas fraction. The weaker accretion rates in the Uchuu simulation, however, are a result of the lower mass resolution of the simulation. In shallow gradient bins, there is a clear flat profile to Uchuu accretion histories, being somewhat reminiscent of \ac{illtng} samples which were discarded by our quality cuts.

With Uchuu being of lower mass resolution than any of the \ac{illtng} simulations, the resolution of low mass haloes will be poor. In the lowest mass quintile, galaxy evolution may be poorly predicted from the start of the galaxy's evolution due to the sensitivity of the star formation algorithm in \ac{illtng} to the number of dark matter particles \citep{Pillepich2018}. Haloes with shallower accretion gradients generally form at later times \citep{Montero-Dorta}, as is the case for our Uchuu data, and therefore those in any given mass bin will be of lower mass at any nonzero redshift. They are therefore subject to a similar resolution effect. For satellites with shallow accretion, there is a noticeable decline in the median accretion rate compared with \ac{illtng}, probably due to the absence of low mass objects being accreted onto the halo.

\subsection{Half-Mass Radius}
\label{sec:rhalf}

\begin{figure*}
\includegraphics[width=\linewidth]{"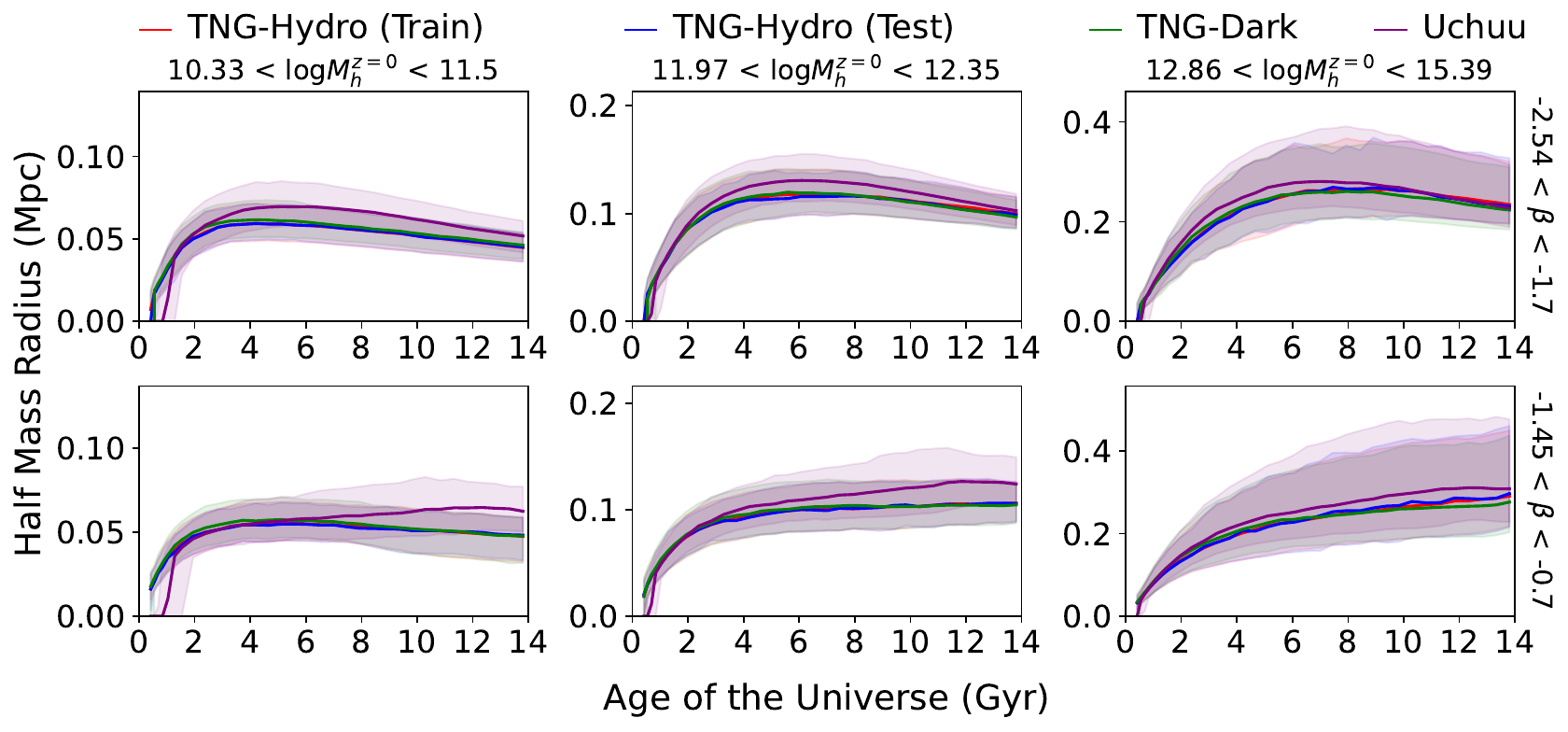"}
\vspace*{-20pt}
\caption{Growth of the half-mass radius of central haloes over time, shown in the same tabular format as \cref{fig:Mhdot}, with the same bins of halo mass and accretion gradient.}
\label{fig:Rhalf}
\end{figure*}

\begin{figure*}
\includegraphics[width=\linewidth]{"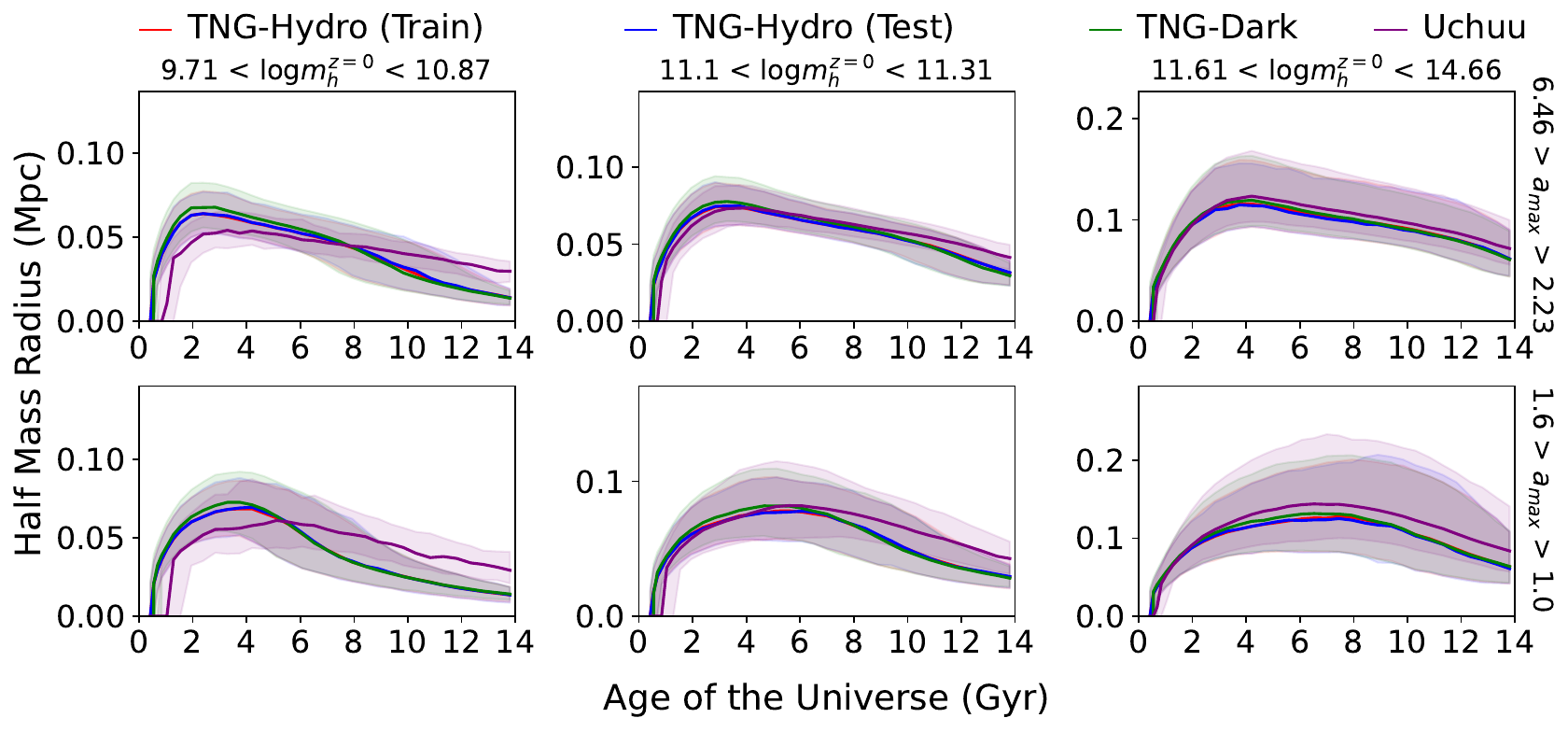"}
\vspace*{-20pt}
\caption{Growth of the half-mass radius of satellite subhaloes over time, shown in the same tabular format as \cref{fig:mhdot}, with the same bins of halo mass and scaled accretion time.}
\label{fig:rhalf}
\end{figure*}

The quantity which differs most starkly in the Uchuu data, however, is the halo half-mass radius, shown for central and satellite haloes in \cref{fig:Rhalf,fig:rhalf} respectively. There is a clear deviation in the size evolution of haloes in Uchuu compared with \ac{illtng} in low mass and shallow gradient bins. However, they are also subtly but noticeably larger in most other bins and at most times.

In TNG-Dark, the half-mass radii of central haloes are subtly smaller in high mass bins than in the other simulations. This may be explained by the results of \citet{Haggar} and \citet{Riggs}, who show that the number density of haloes which are gravitationally bound to a galaxy group or cluster is underpredicted in dark simulations with respect to their hydrodynamical counterparts, within two virial radii of the cluster or galaxy group. This is caused not only by the high density of baryons in the large object's centre but the effect this has on its own density profile.

\citet{Chua_2019,Chua} show that the radial profiles and asymmetries of haloes in \ac{illtng}, and in the original Illustris simulation, are affected significantly by baryons. According to our data, however, this does not affect the size history in TNG-Dark when compared with the full physics simulation. In \citetalias{Chittenden} we thoroughly compared quantities in hydrodynamical and dark \ac{illtng} simulations at all times to ensure that they could be used in a dark matter simulation. In Uchuu, however, the half-mass radius is defined in terms of virial mass, unlike the SubFind result in \ac{illtng}. The SubFind algorithm used in \ac{illtng} defines the boundaries of a subhalo according to a contour of constant density meeting a saddle point in the local density field \citep{Springel_2001}, and the half-mass radius is defined as the radius which encloses half of the mass enclosed by this boundary. In Uchuu, a spherical region whose radius encloses the virial mass is assumed \citep{Uchuu}, and therefore the calculation of half-mass radius is dependent on the halo density profile. The Rockstar halo finder \citep{Rockstar} used in Uchuu calculates this profile from the cumulative binding energies of halo particles, and therefore is sensitive to the NFW profile \citep{NFW} concentration parameter.

\citet{Zhao} show that, in N-body simulations, the increase of the NFW concentration parameter over time is scaled by the time of formation of $4\%$ of the final halo mass; while \citet{Prada} show that halo concentration is additionally sensitive to fluctuations in the linear density field on the scale of the halo's mass. These are both quantities which will depend on the resolution of the simulation, however this effect on the growth of halo concentration is most likely reflected in the more extended mass distributions at late times, as well as the smaller radii of low mass haloes at early times. \citet{Uchuu} also show that in the smaller, higher resolution Shin-Uchuu simulation, while the halo mass functions do not differ significantly between the two Uchuu models, the mass-concentration relation does grow differently between the two simulations, illustrating that this difference in structure is in fact a resolution effect. Morphological halo quantities such as virial velocity and axis ratios are additionally dependent on the gravitational softening scale, which, being larger in Uchuu than in \ac{illtng}, would result in a flatter $M_h - v_\text{max}$ relation at low mass, and thus a similarly distorted mass-concentration relation \citep{Mansfield}.

For both central and satellite haloes, it is the youngest, smallest haloes which are most affected by the difference in halo concentration owing to Uchuu's lower resolution. In each case these haloes are growing from low mass progenitors and thus more likely to suffer from delayed growth in the Uchuu data. Furthermore, \citet{Hoffmann} argue that the subhalo finder algorithm makes little difference to the inferred halo shape, provided that the subhalo is comprised of adequately many particles, which low mass, low resolution haloes do not possess. We therefore stress that the delayed growth of the half-mass radius is a resolution effect and not dependent on the algorithm which defines the halo. Yet on the other hand, \citet{Onions} argue that Rockstar is superior to most algorithms in identifying halo substructures in dense halo centres, which can affect the half-mass radius value. The fact that some graphics in \cref{fig:Rhalf,fig:rhalf} show a larger range of radii in Uchuu for all times, suggests that this is also a relevant factor.

\subsection{Circular Velocity}
\label{sec:vcirc}

The proxy for the virial circular velocity used in our model depends on the above two quantities. It was defined specifically due to extreme differences between the maximum orbital velocities in TNG-Hydro and TNG-Dark, particularly at early times.

Due to the discrepancies we have shown for these quantities in the Uchuu data, the median values for this proxy are slightly too small, but the shape of this median curve remains congruous to its \ac{illtng} counterparts. In low mass bins, however, this proxy is overestimated in TNG-Dark; most likely as a result of excess mass accretion. This also happens to high mass centrals, however to a smaller extent and owing instead to smaller half-mass radii.

\subsection{Cosmic Environment}
\label{sec:env}

In \citetalias{Chittenden} we characterised the environmental histories of central and satellite haloes using subhalo overdensities at each \ac{illtng} snapshot, and devised a radial skewness quantity which characterised the interaction histories of these haloes. While the mass and structure quantities discussed thus far were deemed important to predicting the star formation histories of the haloes' galaxies, these environmental properties were shown to predict their chemical enrichment.

\begin{figure*}
\includegraphics[width=\linewidth]{"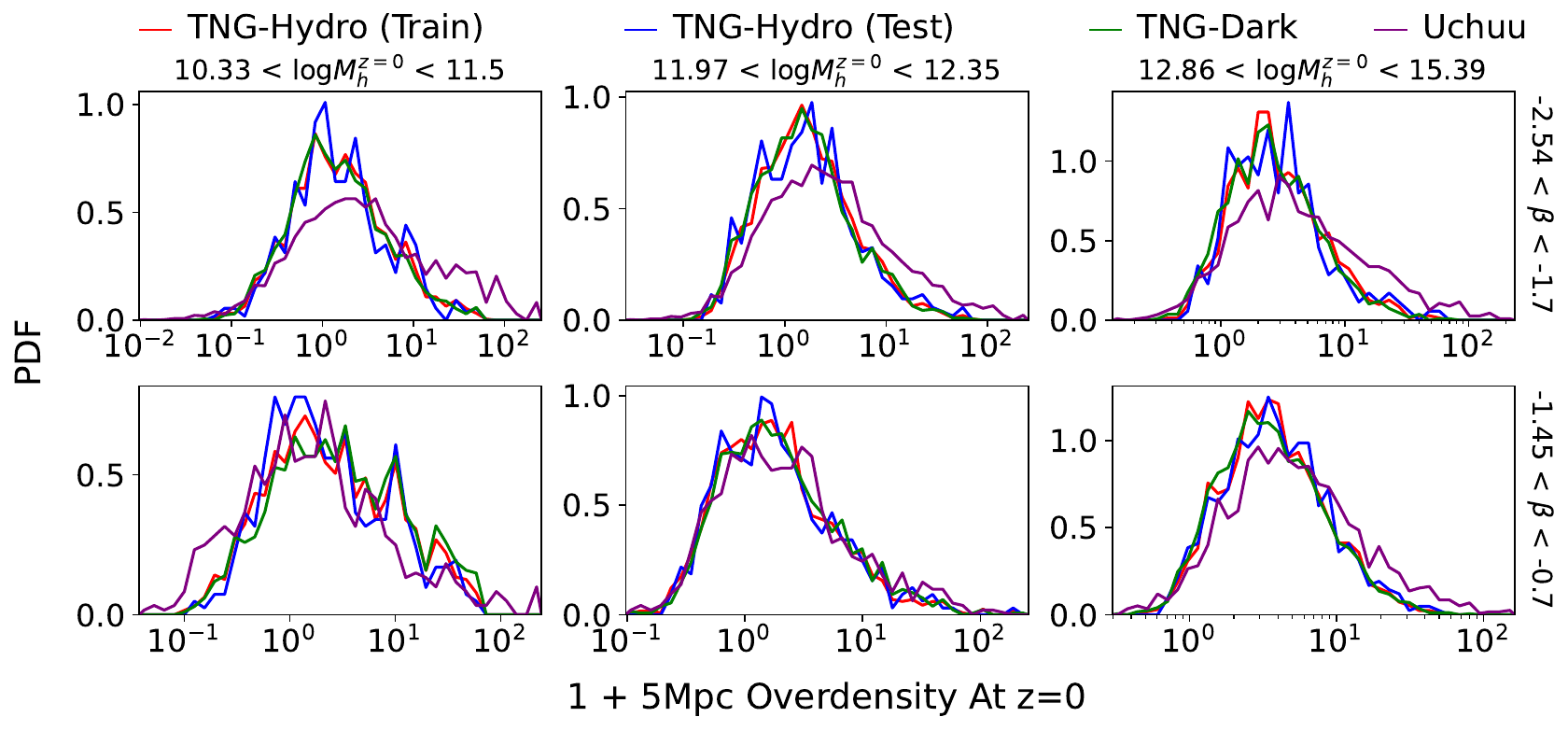"}
\vspace*{-20pt}
\caption{PDFs of the dark matter overdensities enclosed in a 5Mpc spherical region surrounding zero-redshift haloes in the four simulation datasets. These histograms show similar distributions of overdensity for a given mass and accretion history, except for typically larger densities in the Uchuu simulation.}
\label{fig:delta5zlow}
\end{figure*}

As previously discussed, the definitions of halo substructure differ in the Uchuu catalogue, using Rockstar haloes in the place of subhaloes, and thus the overdensity and skew calculations are based on halo tracers. Overdensities in Uchuu are larger than those in TNG-Dark, which can be seen in \cref{fig:delta5zlow}. In a lower resolution simulation, the calculation of environmental quantities is more sensitive to edge effects of the calculation volume, and the radially symmetric definition of a halo in Uchuu can affect the centre of mass of a given object, which may both be contributing factors to the differences in overdensity, possibly adding substantial residual haloes from the limits of the calculation aperture in dense environments. Skews, on the contrary, are not noticeably affected by the resolution difference. As this is a mass-weighted quantity and thus independent of any scaling, it is likely that the difference in overdensities is a result of the mass content of the relevant density tracers.

\section{Galaxy Predictions}
\label{sec:pred}

Having outlined the causes of differences between haloes and environments in each simulation, we explain in this section how the implementation of these distinct variables into the neural network changes important results. We discuss the effects that the simulation differences have on the direct predictions and derived halo-galaxy relations, providing a physical explanation into differences in the results. For a detailed quantitative analysis of these results, where we compare Kolmogorov-Smirnov (KS) statistic values relating our fiducial and modified predictions of physical values to the original TNG simulation data, see \cref{sec:quant-analysis}.

\subsection{Star Formation History}
\label{sec:sfh}

\begin{figure*}
\includegraphics[width=\linewidth]{"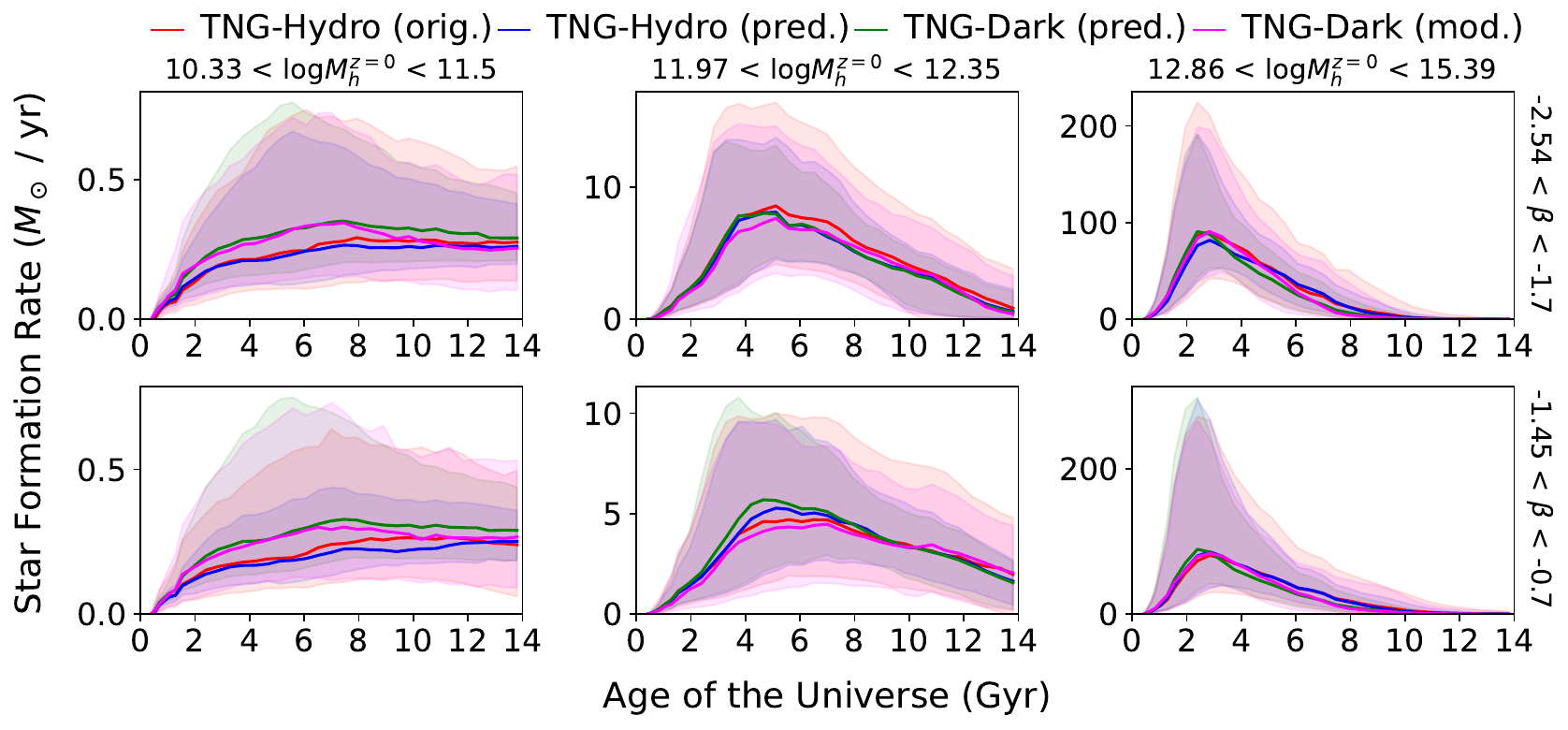"}
\includegraphics[width=\linewidth]{"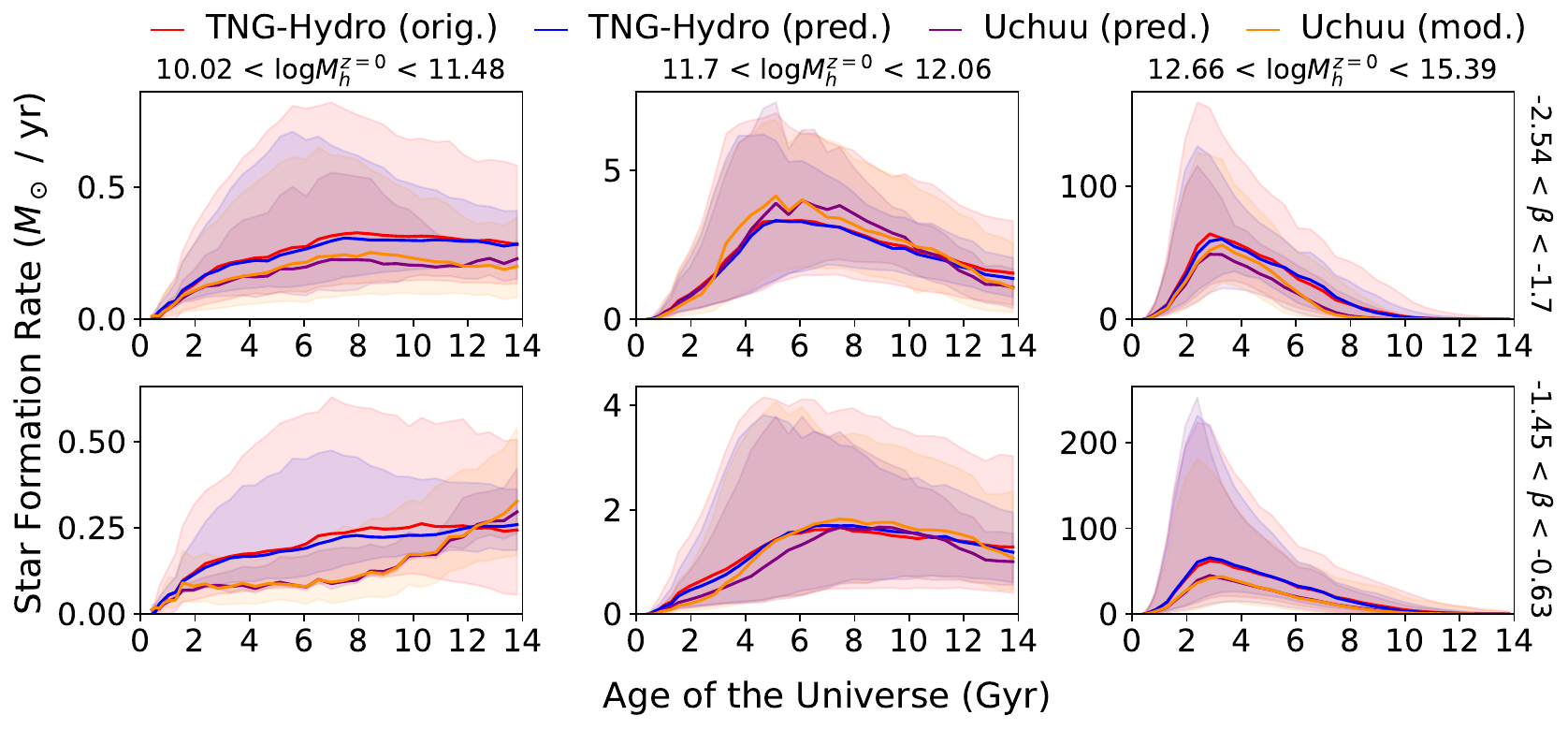"}
\vspace*{-20pt}
\caption{Star formation histories of central galaxies in bins of halo mass and specific mass accretion gradient. In low mass bins, there is an overprediction of star formation rates in TNG-Dark, and an under-prediction in Uchuu. In higher mass bins, this difference becomes smaller. The stochastic modification has shown here to reshape the shape and variance of N-body predictions to more closely resemble the orginal TNG data, improving upon the predictions in the hydrodynamical simulation as well. However, the modification cannot amend the poorly predicted star formation histories in low mass objects, as the predicted Fourier Transforms used by the modification are similarly affected by mass resolution.}
\label{fig:Sfh}
\end{figure*}

\begin{figure*}
\includegraphics[width=\linewidth]{"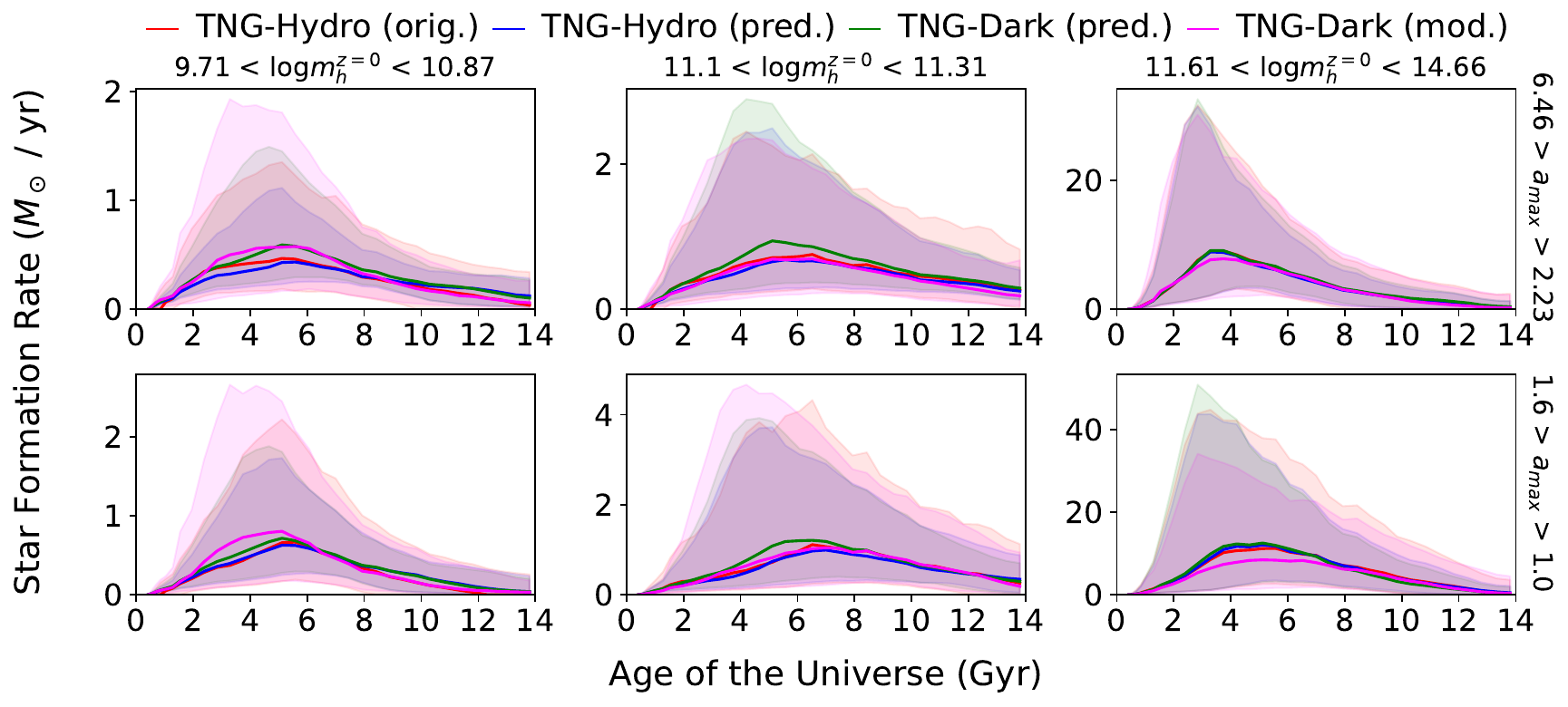"}
\includegraphics[width=\linewidth]{"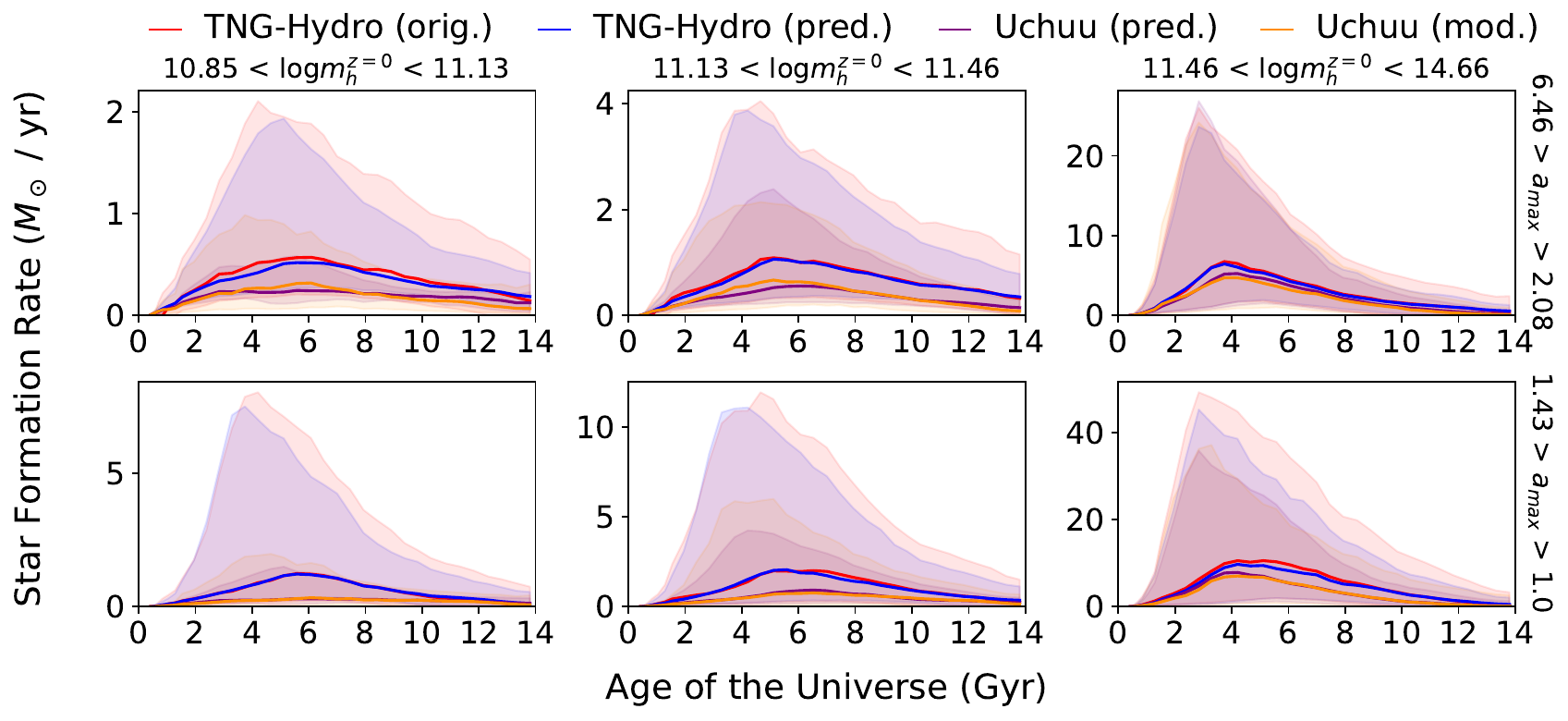"}
\vspace*{-20pt}
\caption{Star formation histories of satellite galaxies, binned in the same manner as in \cref{fig:Sfh}. This figure shows apparently poor predictions of star formation histories of low mass galaxies in Uchuu, however many of these bins are underpopulated by the Uchuu data and contain low quality haloes.}
\label{fig:sfh}
\end{figure*}

\Cref{fig:Sfh,fig:sfh} show the median and $15^\text{th}-85^\text{th}$ percentile ranges of predicted star formation histories in bins of final (sub)halo mass and mass accretion gradient, including the modified star formation histories in the N-body simulations. Each of these figures show that galaxies in high mass and fast accretion bins are typically well matched, but in low mass bins the differences are stark. Star formation rates in TNG-Dark are over-predicted, while in Uchuu they are underpredicted.

It may appear that the majority of Uchuu galaxies, particularly satellites in \cref{fig:sfh}, are severely underpredicted in their stellar mass. This is misleading due to the absence of low mass haloes, discussed in \cref{sec:uchuures}. The two lowest mass quintiles of \ac{illtng} haloes contain $\sim 12 \%$ of satellite galaxies in Uchuu. However, these star formation histories are nonetheless severely underpredicted. Their halo histories have already shown that these objects accumulate their mass later and more slowly than their \ac{illtng} counterparts. This can be seen clearly for satellites in \cref{fig:mhdot,fig:rhalf}.

In \cref{tab:corr} we show Spearman correlation coefficients between halo and galaxy properties in a narrow, low mass bin for central and satellite objects in Uchuu. Within individual bins at low mass, the final stellar mass is correlated strongly with the proxy for circular velocity, which in the Uchuu data is subtly smaller than its \ac{illtng} equivalents. Weaker correlations exist with stellar mass and half-mass radius and overdensity, which are also somewhat different in Uchuu.

However, particularly for low mass satellites in \cref{fig:sfh}, underpredicted star formation histories are clearly correlated with similarly undermined mass accretion histories in \cref{fig:mhdot}, which will have had a causal effect on their galaxy growth. Mass accretion at early times is particularly important for the acquisition of star-forming gas, which would explain the lack of subsequent star formation in these predictions. In the \ac{illtng} simulations, low-mass haloes are particularly gas-rich \citep{JJDavies2020} and so the star formation at early times is likely to be sensitive to the lack of accumulation of low mass progenitors.

\begin{table}
\begin{center}
\begin{tabular}{|c|c|c|c|c|}
\hline
 \multicolumn{1}{|c|}{\multirow{2}{*}{Input Variable}} & \multicolumn{2}{c|}{Stellar Mass} & \multicolumn{2}{c|}{Metallicity} \\ \cline{2-5}
 & Central & Satellite & Central & Satellite \\ \hline \hline
 Circular Velocity & 0.783 & 0.637 & 0.693 & 0.361 \\ \hline
 Half-Mass Radius & -0.566 & -0.573 & -0.602 & -0.314 \\ \hline
 Overdensity & 0.176 & 0.298 & 0.174 & 0.142 \\ \hline
\end{tabular}
\end{center}
\caption{Spearman correlation coefficients of multiple halo properties with stellar mass and metallicity, in narrow, low halo mass bins. All quantities shown here are considered at $z=0$ in the Uchuu simulation. We evaluate these within halo mass ranges of $11.77 < \log_{10} M_h^{z=0} < 11.97$ for centrals, and $10.9 < \log_{10} m_h^{z=0} < 11.1$ for satellites.}
\label{tab:corr}
\end{table}

Comparing the network's performance on TNG-Dark with TNG-Hydro predictions, we see that these star formation histories are enhanced rather than suppressed, which correlates with similarly excessive mass accretion rates. \citet{Sorini} show that gas accretion and stellar feedback processes have their greatest influence on the size and shape of haloes and large scale structure at higher redshifts, and that stellar feedback is the principal cause of suppression of the star formation of low mass objects. The lack of stellar feedback in dark simulations will have limited the effects which result in halo mass loss, which are particularly prominent effects for low mass objects. The excess mass accretion in TNG-Dark will have resulted in excess star formation being predicted.

\begin{figure}
\includegraphics[width=\linewidth]{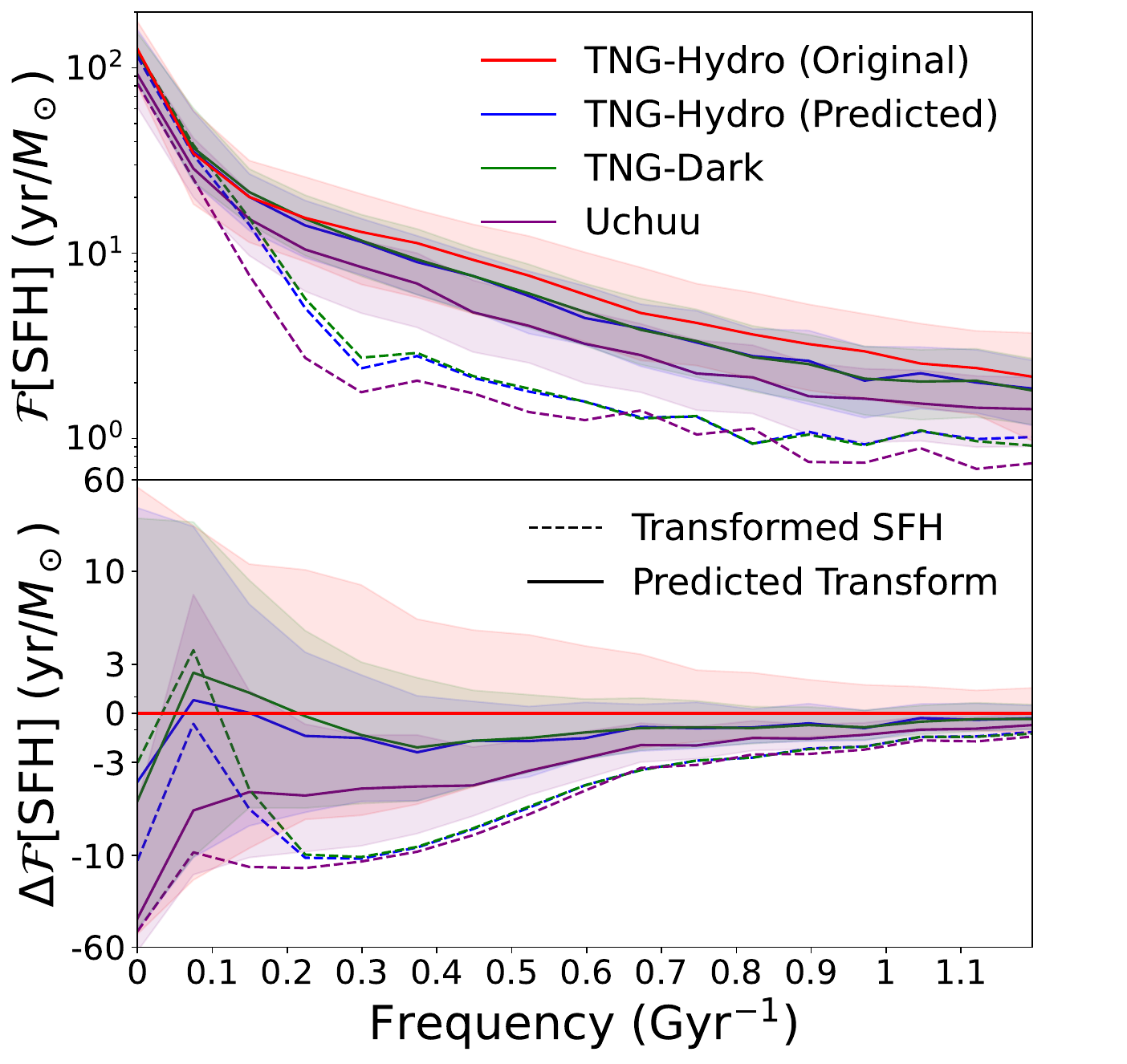}
\vspace*{-20pt}
\caption{A figure comparing the median and interquartile ranges of predicted Fourier Transforms of star formation histories, for central galaxies where $11.97<\text{log}_{10} M_h^{z=0}<12.35$ and $-1.45<\beta<-0.7$, corresponding to the centre-bottom panel in \cref{fig:Sfh}. The top panel of this figure shows the Fourier Transforms themselves, while the bottom panel shows the difference in the Fourier amplitude with respect to the median target amplitude. Where applicable, dashed and solid lines indicate the Fourier Transforms of the star formation histories predicted by the neural network, and the predicted Fourier Transforms when the network is trained to predict these, respectively. This illustrates that for all datasets, the median Fourier Transforms are all of similar shape to the target TNG-Hydro data (red) when predicted by the neural network. The TNG-Dark result (green) is similar enough to the TNG-Hydro predictions (blue) that the modified data in TNG-Dark will be just as accurate as in our companion paper. On the contrary, the Uchuu data (purple) has clearly suppressed amplitudes at all frequencies, and does not align with the target distribution. This suggests that star formation modes which influence our results will be absent from some of our Uchuu predictions, regardless of the power of the stochastic amendment.}
\label{fig:fsfhdiff}
\end{figure}

The stochastic modification to our star formation and metallicity histories, explained in detail in \citetalias{Behera}, introduces fluctuations to individual galaxies by drawing random phases from high frequency modes, which are scaled in amplitude according to a predicted Fourier Transform of the sample's \ac{sfh} and \ac{zh}. By training the neural network introduced in \citetalias{Chittenden} to predict the Fourier amplitude of each \ac{sfh} and \ac{zh}, while changing none of the input variables or aspects of the network's design, we predict Fourier amplitudes which resemble those of the original \ac{illtng} data; unlike the fiducial predictions of the original network, which lose much of this information at high frequencies. A comparison of the Fourier Transforms derived from central star formation histories in an intermediate mass bin is shown in \cref{fig:fsfhdiff}.

With the stochastic modification, the additional fluctuations in the star formation histories have recovered enough information that the median and variance of the predictions in both N-body simulations are improved, with the dataset more akin to the target \ac{illtng} data. However, the correction is less effective for low mass centrals and satellites, and does not resolve the severe offsets seen in Uchuu galaxies.

As star formation histories are poorly predicted for under-resolved Uchuu galaxies, so are their Fourier Transforms, making mass resolution crucial to the stochastic modification as much as the machine learning model itself. We illustrate this in \cref{fig:fsfhdiff}, comparing the predicted Fourier amplitudes of each simulation with the desired result from the raw \ac{illtng} data. For this bin in $\beta$ and $M_h$, we can see that the Uchuu Fourier amplitudes are reduced across all frequencies. With the stochastic method sampling high frequency modes with suppressed amplitudes, the same variable star formation history will be lacking in under-resolved Uchuu data, regardless of the efficacy of the stochastic method.

In \citetalias{Behera}, several post-processing filters were applied to mitigate unphysical results in the modified data. For Uchuu and TNG-Dark, an additional threshold is necessary due to unphysical distortions arising from less accurate predictions of the Fourier Transform amplitudes. To address this, we impose a threshold that limits the relative difference between the corrections and the predictions to less than a factor of 1.5, which is typical of the modified TNG-Hydro data. Any larger deviations are replaced by the fiducial predictions to avoid further distortions.

For the TNG-Dark data in this mass/gradient range, the Fourier Transforms resemble the TNG-Hydro predictions, suggesting that the same stochastic data as in \citetalias{Behera} may appear in TNG-Dark, yet as discussed below, the modifications in TNG-Dark are also subject to certain resolution effects.

\begin{figure*}
\includegraphics[width=\linewidth]{"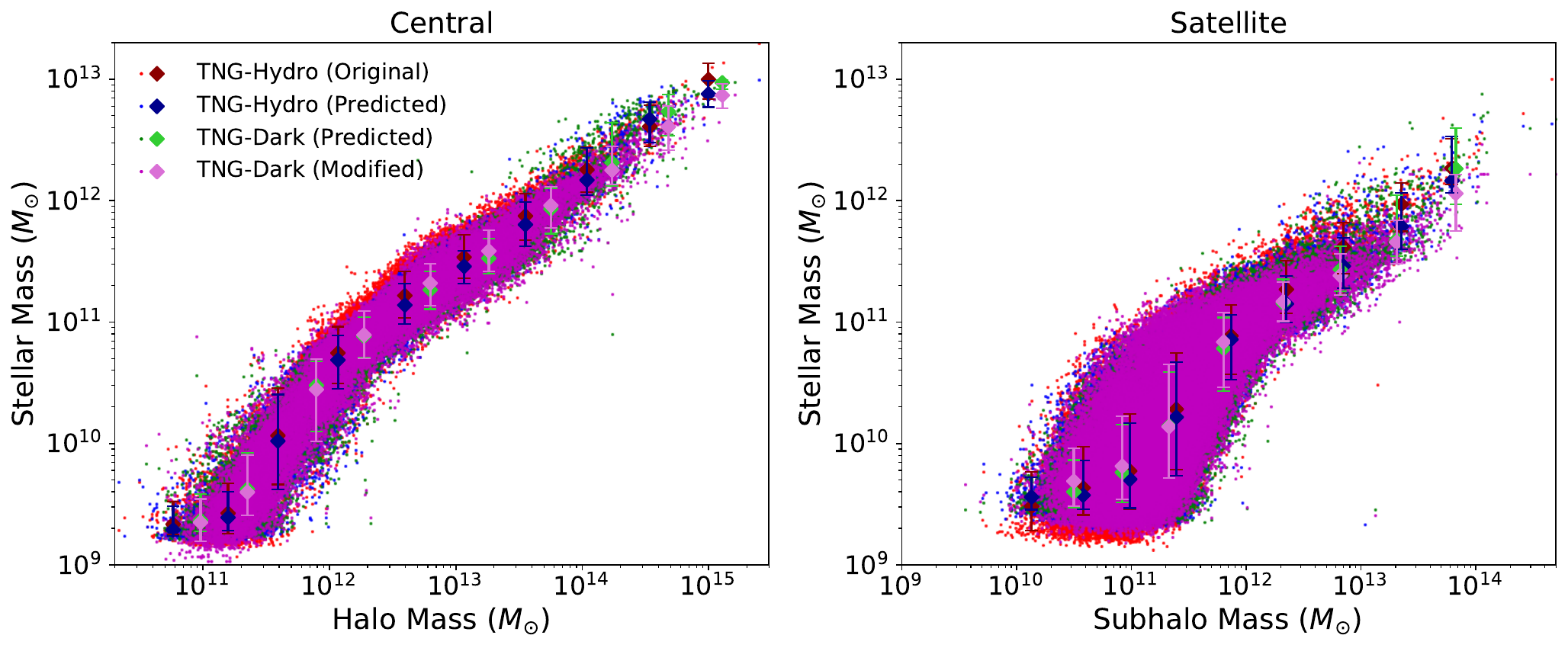"}
\includegraphics[width=\linewidth]{"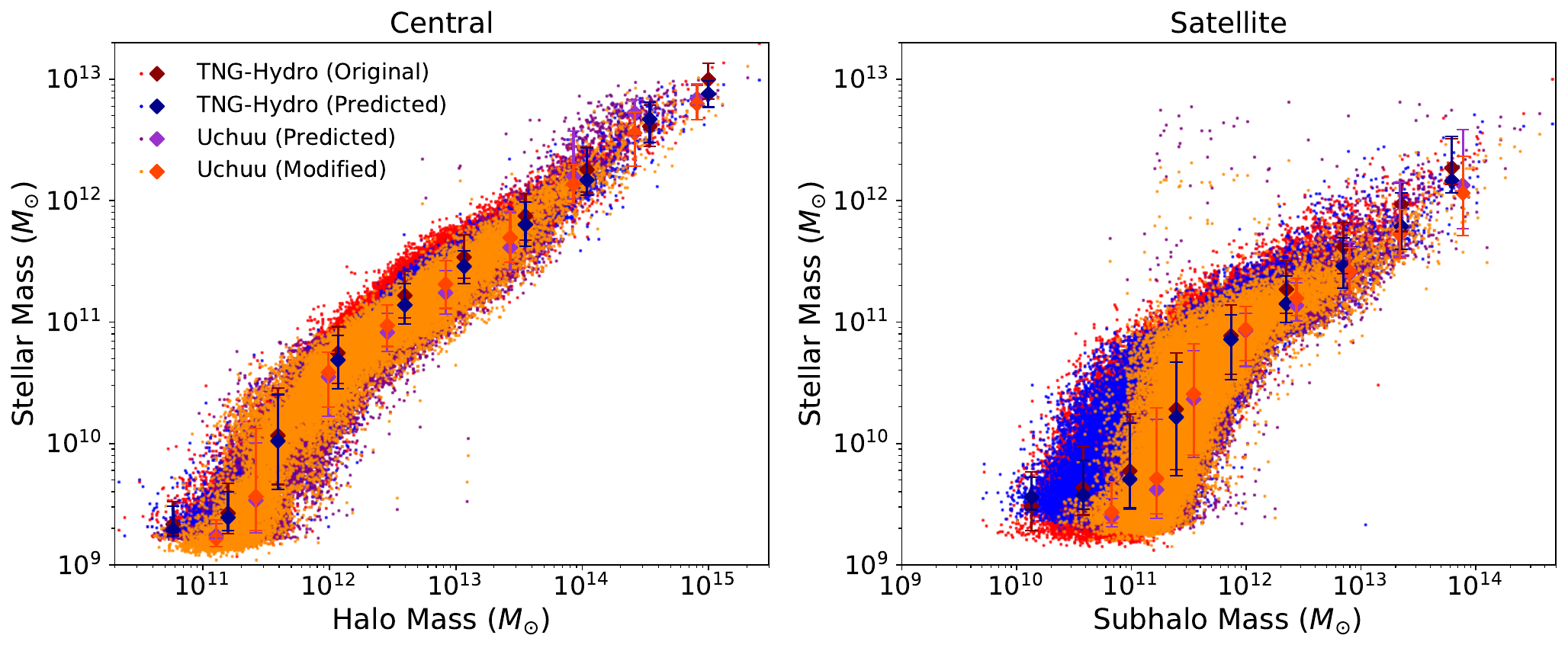"}
\vspace*{-20pt}
\caption{The graph displays the numerical \ac{shmr} of central galaxies (left) and satellite galaxies (right) based on their star formation rates. Each galaxy is represented by a data point, while the error bars indicate the median and $15^\text{th}$ and $85^\text{th}$ percentiles of stellar mass within a specific halo mass range. The comparable shapes of these relationships suggest an accurate prediction of the star formation histories.}
\label{fig:shmr}
\end{figure*}

Numerically integrating the \ac{sfh}s of each simulation result in self-consistent and accurate \ac{shmr}s, shown in \cref{fig:shmr}. Between predictions in the TNG-Hydro and TNG-Dark simulations, there is little noticeable difference between the relations, other than slightly reduced scatter in the dark predictions. In \cref{sec:KS1}, we show that the halo mass-binned stellar masses and scatters differ on average by around 0.01-0.02 dex following modification. For Uchuu, the difference in stellar and halo mass distributions is apparent, and this is most profoundly so at low masses, with satellite KS statistic values of $\sim 0.4$ even after modification. From intermediate to high masses, there is a slight underprediction of mass (0.05-0.15 dex) and an offset in scatter (0.03-0.05 dex), but the Uchuu \ac{shmr} remains very well matched to the predictions in TNG-Hydro and TNG-Dark, indicated by KS statistics in the approximate range of 0.1-0.2.

In both the central and satellite \ac{shmr}s, particularly at intermediate masses, the stochastic modification makes subtle improvements to the average height of the relation, as well as marginally increasing the scatter. We show a closer fit to the target distributions of stellar mass following modification in \cref{sec:KS1}, where the KS statistics for halo mass bins above $10^{12} M_\odot$ are reduced by a median of $\sim 37.7\%$. The fluctuations added by the modification add a component of stellar mass which may be attributed to events such as galactic winds or minor mergers; while the cycling of gas through the CGM acts on sub-Gyr timescales for high-mass galaxies in dense environments \citep{Oppenheimer}, stellar feedback is more likely to disrupt the shallow potentials of low mass galaxies and contribute to greater mass loss on short timescales \citep{ElBadry, AA}. Despite the variety of processes acting on different timescales in different mass regimes, the fluctuations implemented by the modification are rarer for the most massive galaxies, and smaller relative to the total mass, making little change to the modified stellar mass.

At low masses, we see that the distribution of stellar masses in Uchuu is significantly biased towards low masses, owing to the compromised star formation rates. Low mass satellite haloes have already been discarded by our cuts, resulting in a clearly distinct halo mass distribution, yet there are still a large number of objects below $\sim  2 \times 10^{11} M_\odot$ with severely miscalculated stellar mass. Satellite objects within this mass range clearly cannot be trusted.

Low mass satellites are additionally made impractical due to a limitation of the resolution correction presented in \citetalias{Chittenden}. The correction consists of a $z=0$ halo mass dependent ratio between star formation and metallicity histories in TNG100-1 and the lower resolution run TNG100-2, with a mass resolution consistent with TNG300-1; to adjust TNG300-1 galaxies to match TNG100-1 data prior to their simultaneous use as training data. For satellite galaxies, however, the correction ceases to be accurate when satellite galaxies of distinct evolutionary geometries (e.g. early-forming, quenched and late-forming, star-forming galaxies), are combined as one correction for all low mass satellites; as well as when the TNG100-2 galaxies on which the correction is based are affected by low sample size and resolution.

The correction was based on a method used by \citet{Pillepich2018}, who find that the radius enclosing star particles influenced their corrected data. As \citet{IllustrisTNG} find higher gas densities in TNG300-1 than in TNG100-2, it is possible that differences in the density of gas surrounding satellites results in differences in ram pressure stripping and thus stellar mass profiles. These are two effects which reduce the validity of the correction for satellites, particularly at low mass. It is such a substantial difference for low mass satellites that these samples are ignored when applying the stochastic correction.

\begin{figure}
\includegraphics[width=\linewidth]{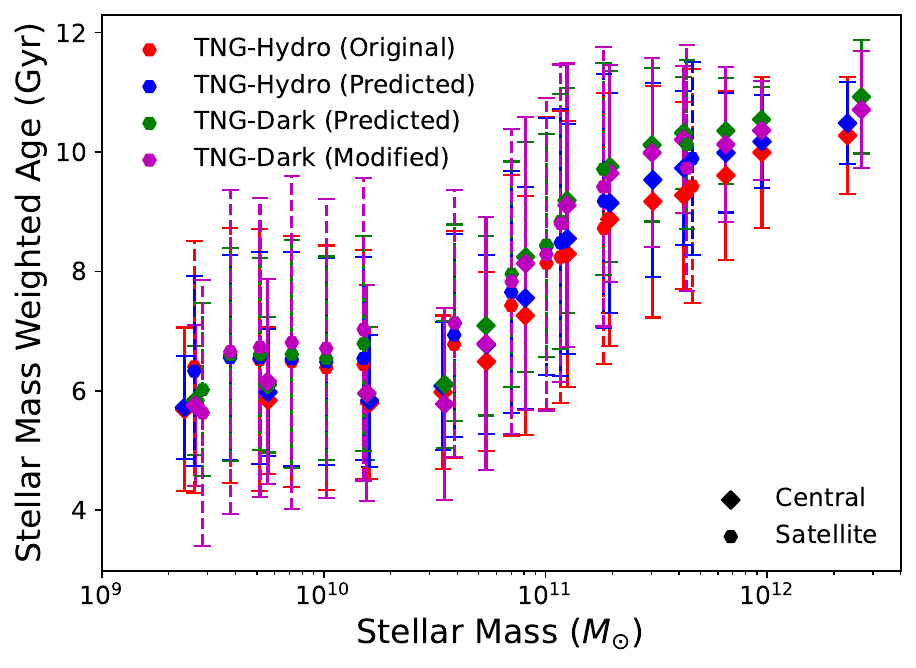}
\includegraphics[width=\linewidth]{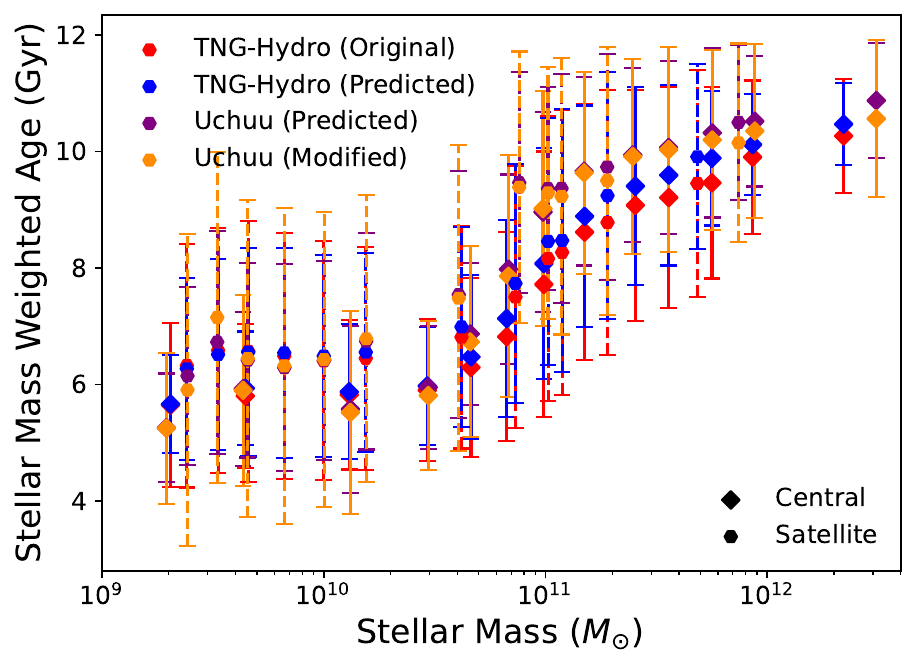}
\vspace*{-20pt}
\caption{The figure depicts the mass-weighted ages of central and satellite galaxies, showing TNG-Dark results in the top panel and Uchuu results in the bottom panel. The ages are plotted against the predicted stellar mass, with the median and interquartile range of ages shown in various mass bins. The plot reveals a precise recovery of the age-mass trend in the dark simulations. There is a bias towards higher ages in the dark simulations which grows with mass, partially corrected by the stochastic modification.}
\label{fig:mwa}
\end{figure}

In high mass bins, the majority of stellar mass is formed earlier than in TNG-Hydro, for both dark simulation suites. This is illustrated by \cref{fig:mwa}, which shows mass weighted ages of galaxies in bins of stellar mass, and shows that high mass galaxies are biased towards older ages. In part, this can be seen by a sharp increase in both halo and stellar mass at earlier times, but in \cref{fig:Sfh,fig:sfh} the following decline in star formation rate happens sooner. The stochastic modification corrects for excess early star formation, indicating that the missing stellar feedback effect is encoded in the SFH power spectra; yet the modification does not influence the quenching tail, which is a smoother feature and so potentially unseen by the modification due to frequency-dependent phase selection, while unphysical noise in the quenching tail is eliminated by our post-processing filters (see \citetalias{Behera} sec. 3.3). In Uchuu, the mass-weighted ages of intermediate mass galaxies are shifted by a lesser degree than in TNG-Dark, potentially due to differences in the accuracy of Fourier amplitudes. For TNG-Dark, star formation histories are initially aligned with TNG-Hydro, but these star formation rates decline and line up with the lower \ac{sfh} in Uchuu. The quenching of these galaxies is therefore more efficient than their hydrodynamical equivalents.

We have shown in \cref{fig:Rhalf} that the half-mass radius of high-mass haloes is underestimated in TNG-Dark, leading to overestimated halo concentration. Since velocity dispersion traces the dynamics of the central halo on sub-kiloparsec scales, this overestimated concentration implies a denser central region. In galaxies of this mass, such central densities are typically dominated by an AGN. As a result, overpredicting halo concentration can lead to overestimates of AGN feedback strength, causing the neural network to quench star formation prematurely. This may be supported by \citet{JJDavies2020}, who show that the expulsion of the circumgalactic medium in \ac{illtng} and the Eagle simulations \citep{Eagle} is correlated strongly with the central black hole mass of the galaxy, which influence the subsequent specific star formation rate, and act on timescales of multiple gigayears \citep{Iyer, Zinger, Walters}. \citet{Bluck} show that central velocity dispersion, effectively measuring the AGN mass, is a critical parameter for galaxy quenching in MANGA observations, which correlates with the circular velocity and half-mass radius of the halo. In \ac{illtng}, these internal feedback quenching mechanisms dominating central galaxy quenching, alongside environmental effects dominating satellite quenching, are shown to qualitatively agree with other hydrodynamical and semi-analytic models, and with low-redshift SDSS data \citep{Donnari1, Donnari2}.

\subsection{Metallicity History}
\label{sec:zh}

\begin{figure*}
\includegraphics[width=\linewidth]{"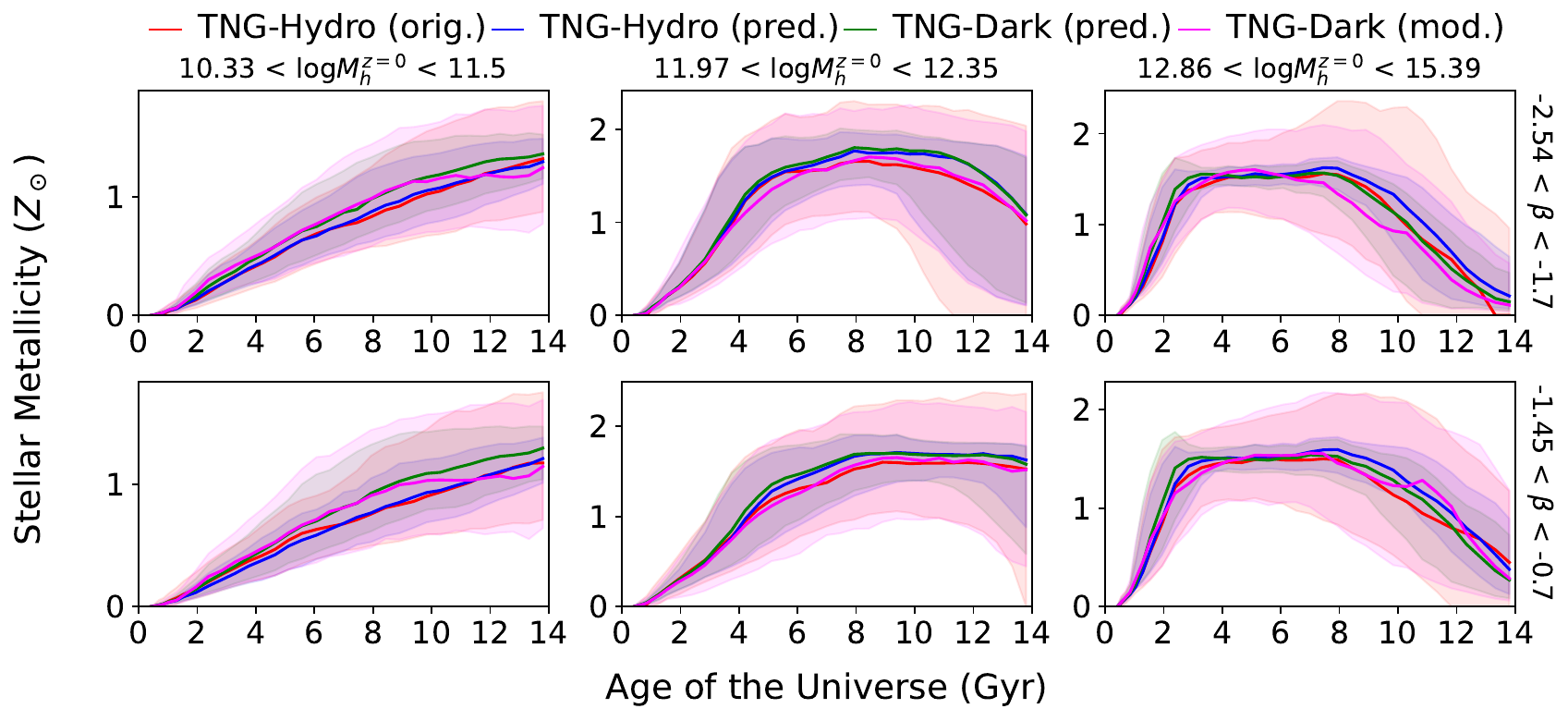"}
\includegraphics[width=\linewidth]{"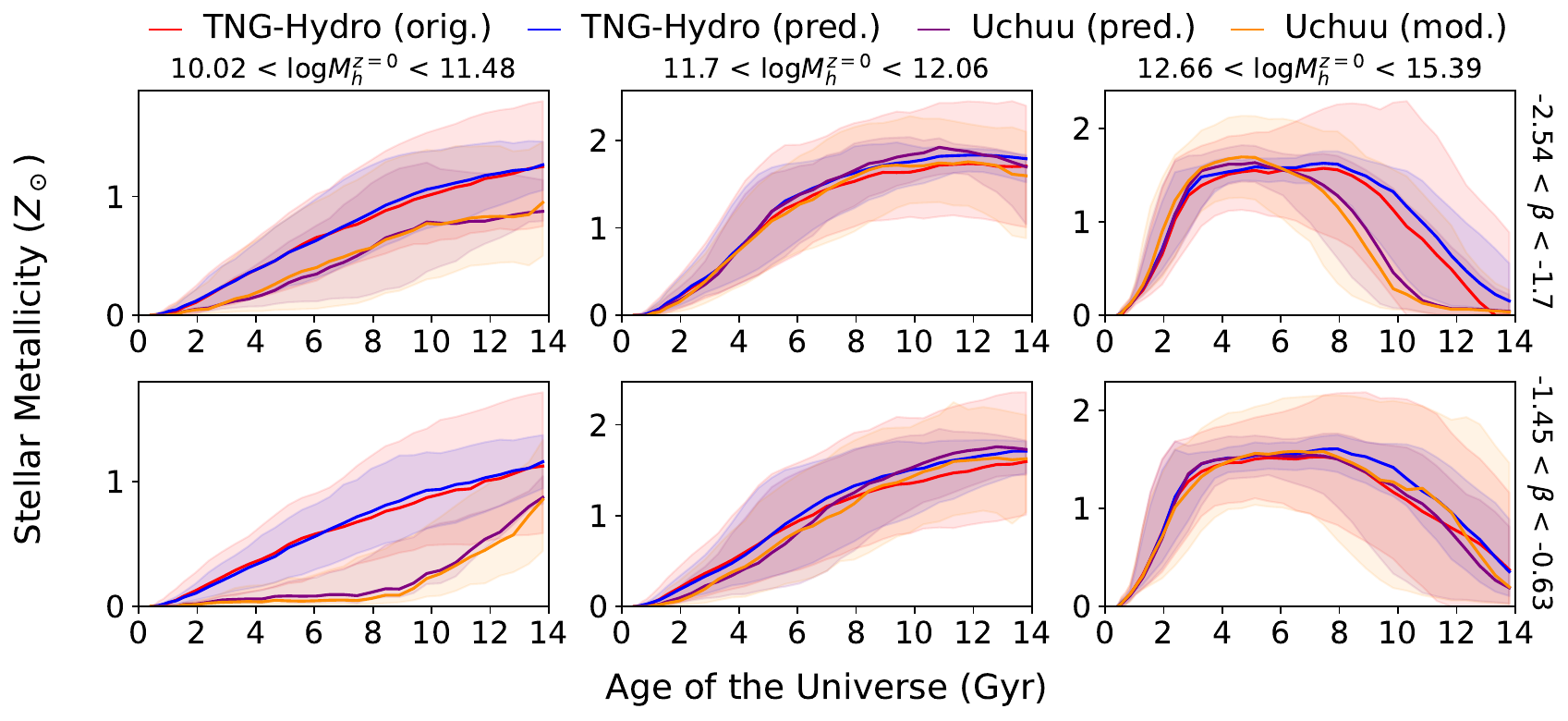"}
\vspace*{-20pt}
\caption{Stellar metallicity histories of central galaxies in bins of halo mass and specific mass accretion gradient.}
\label{fig:Zh}
\end{figure*}

\begin{figure*}
\includegraphics[width=\linewidth]{"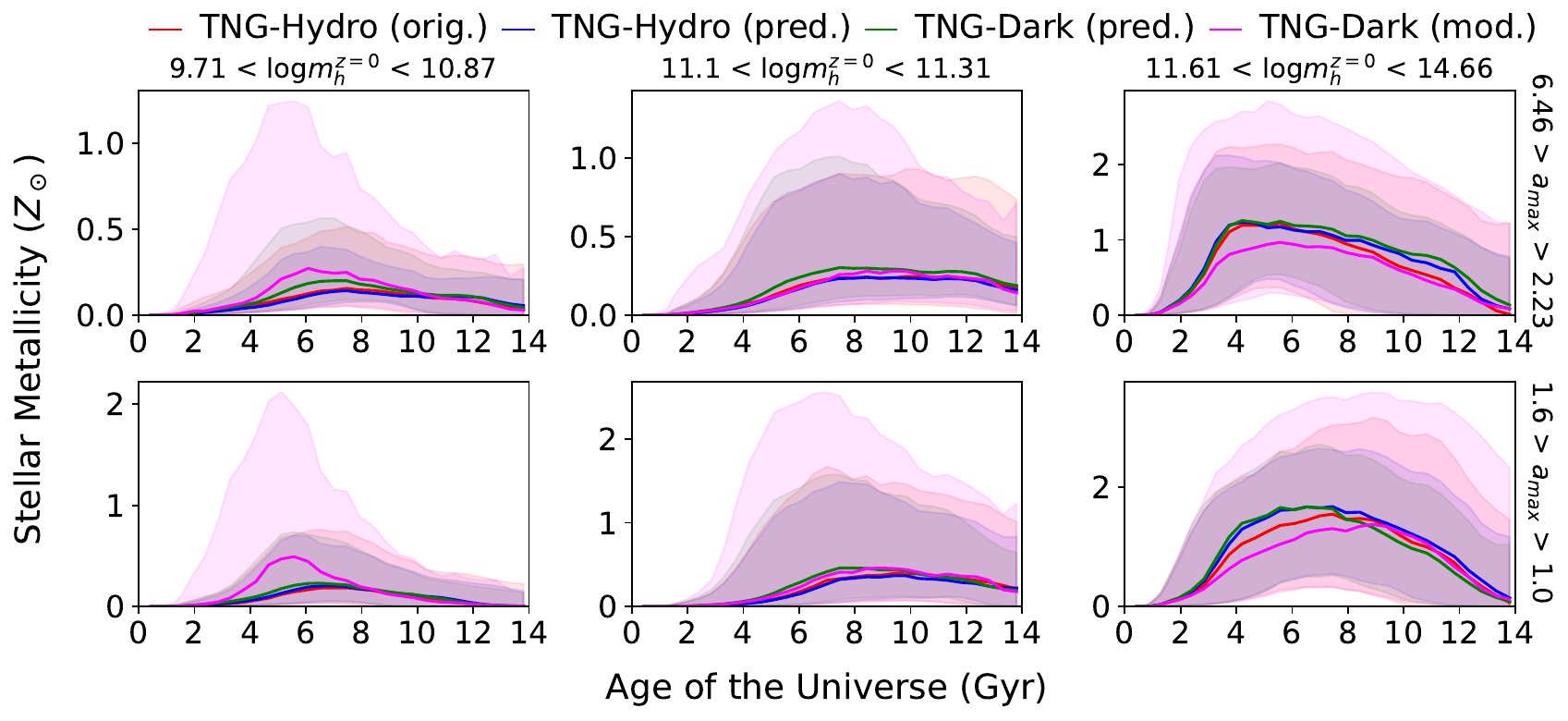"}
\includegraphics[width=\linewidth]{"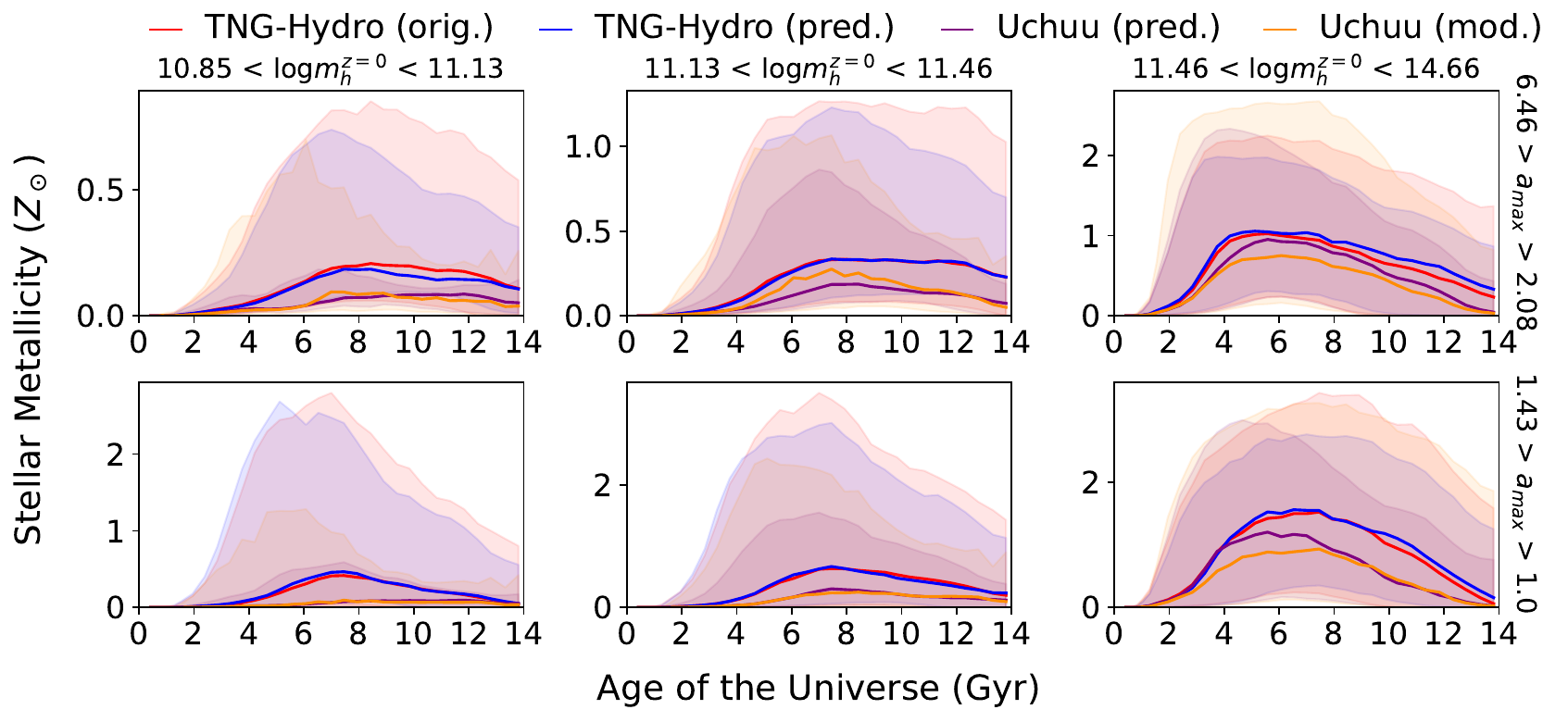"}
\vspace*{-20pt}
\caption{Metallicity histories of satellite galaxies, binned in the same manner as in \cref{fig:Zh}.}
\label{fig:zh}
\end{figure*}

\Cref{fig:Zh,fig:zh} show the median and $15^\text{th}-85^\text{th}$ percentile ranges of predicted metallicity histories. These show a clear failure of the model to predict the metallicity histories of low mass objects in Uchuu, as well as underpredicted results in most satellite galaxies. There is nonetheless a good agreement between Uchuu and \ac{illtng} with most intermediate to high mass central haloes.

The stochastic amendment introduces a significant improvement to the median and variance of metallicity histories, effectively recovering their distribution across mass regimes by 40-70\%, as shown in \cref{sec:KS1}. Like the star formation histories, it is likely that the network-predicted Fourier Transforms contain information on the frequency of chemical enrichment events on short timescales, such as merger-driven star formation. The simultaneous recovery of the variability in star formation and metallicity histories results in a more realistic metallicity distribution with respect to the fiducial network predictions, as shown in \citetalias{Behera}.

We see similar characteristics when comparing metallicity histories seen in TNG-Hydro, TNG-Dark and Uchuu to what we showed for star formation histories in \cref{sec:sfh}. The suppression of metal synthesis in low mass galaxies can be reconciled with the lack of early accretion, particularly as the gas and metal content of these objects' progenitors play an important role in early metal synthesis. The enhanced metal synthesis in low mass TNG-Dark galaxies may also be explained by the abscence of stellar feedback, where stars retain more of their mass and thus produce metals more efficiently.

In narrow mass bins, we find similar but weaker correlations between stellar metallicity and structural and density quantities. The results from \citetalias{Chittenden}, however, suggest that environmental history influences chemical enrichment over time. We show in \cref{fig:delta5zlow} that calculated overdensities are marginally larger in Uchuu, which can influence metallicities by predicting an overabundance of mergers and flybys which redistribute the metals into high mass galaxies, as well as contribute to quenching. The anisotropic nature of these interactions are traced by radial skews, which we show in \citetalias{Chittenden} to have a profound effect on the metallicity history. Though these skews are not well correlated with other halo properties, being difficult to compare between simulations, the low number density of haloes around low mass objects can fail to produce the extreme skews experienced during close interactions which contribute significantly to chemical enrichment.

\begin{figure*}
\includegraphics[width=\linewidth]{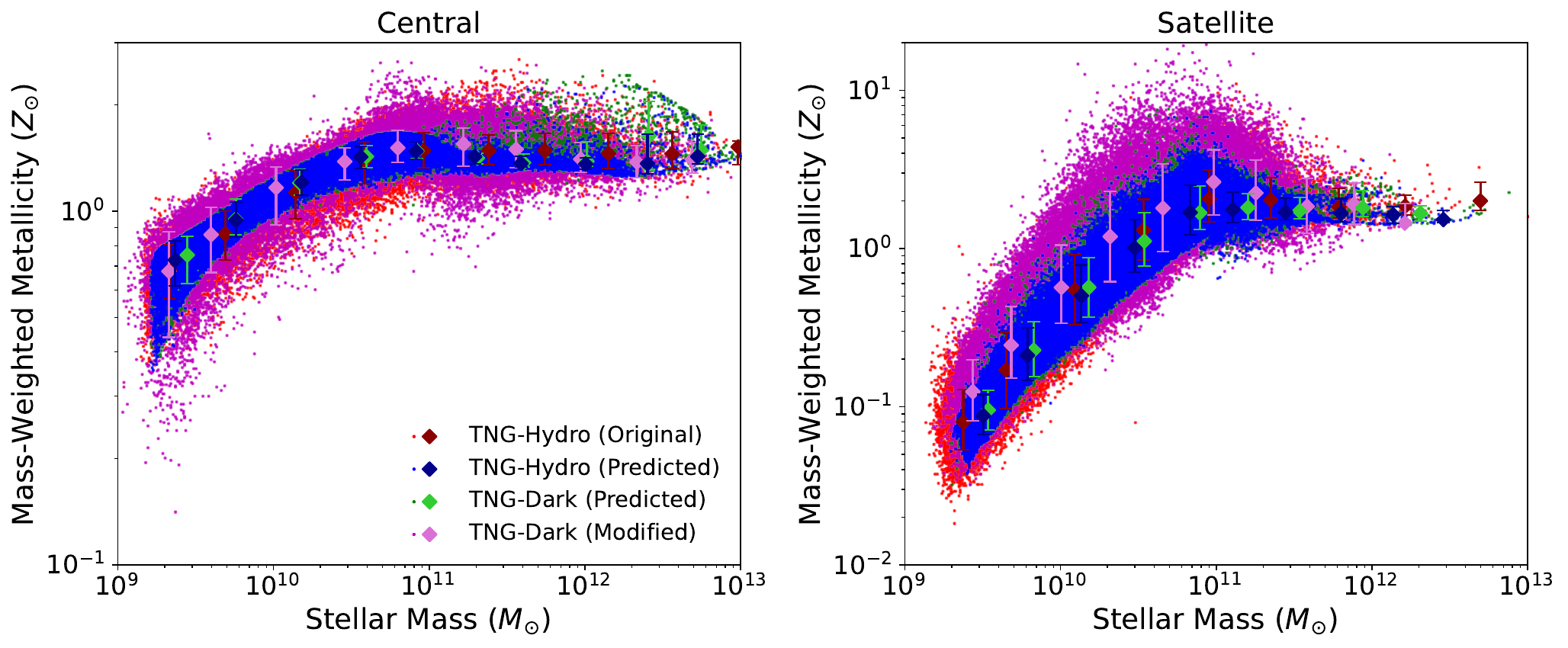}
\includegraphics[width=\linewidth]{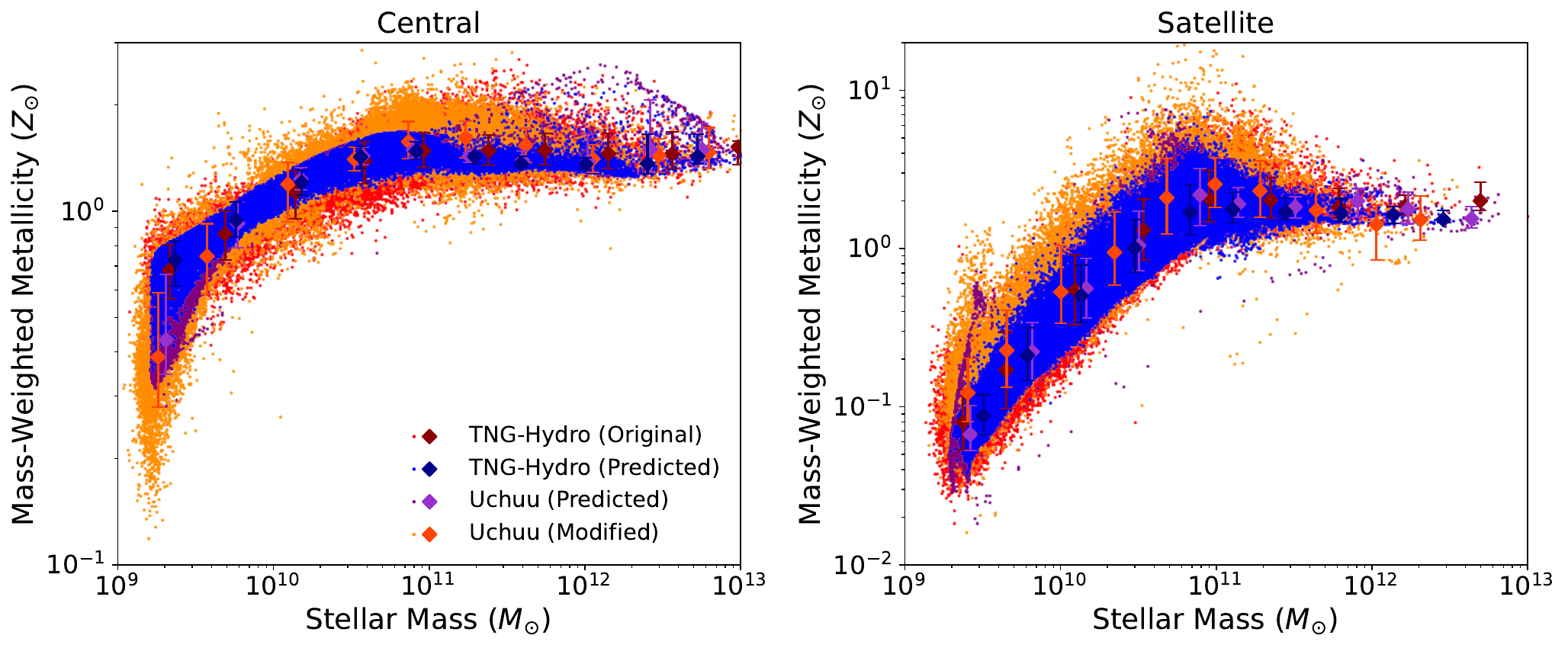}
\vspace*{-20pt}
\caption{Numerical \ac{mzr} of central galaxies (left) and satellite galaxies (right), with numerically evaluated total stellar mass and mass-weighted metallicity.}
\label{fig:mzr}
\end{figure*}

Evaluating the mass-metallicity relations from these star formation and metallicity histories in \cref{fig:mzr}, we see that the underpredicted metallicity histories distort both relations at the low mass end. Nevertheless, the shape and scatter of the relations in the dark simulation are very similar to the network's original TNG-Hydro predictions; particularly following stochastic modification. For central galaxies, however, there are a handful of overpredicted metallicities in the dark simulations at high mass. A Spearman coefficient of 0.683 between age and metallicity for Uchuu galaxies above a stellar mass of $10^{11.5} M_\odot$ shows that these are the same galaxies whose mass-weighted ages are overpredicted. Therefore, there is a greater contribution of metallicity histories at early times to these results, and negligible contribution from when the galaxies begin to lose their star formation. This effect is subsequently mitigated by the stochastic amendment, reducing the coefficient to 0.135; similar to the TNG-Hydro target value of 0.217. As previously established, the modification shifts the early star formation towards the \ac{illtng} target; but additionally adds fluctuations which enhance the chemical evolution across all times. This improvement, not only to the stellar formation time, but to the characteristic time of metal synthesis, is of great value for refining the predicted spectroscopy which would be used in N-body mocks. We discuss these results in the following section.

\section{Mock Observables}
\label{sec:obs}

This section concerns how the previously described similarities and discrepancies between halo and galaxy properties affects the derived observational results, and thus is an assessment of the quality of hypothetical mock survey statistics. As with \cref{sec:pred}, a thorough quantitative analysis supporting these results is shown in \cref{sec:quant-analysis}.

\subsection{Galaxy Spectra}
\label{sec:spec}

\begin{figure*}
\includegraphics[width=\linewidth]{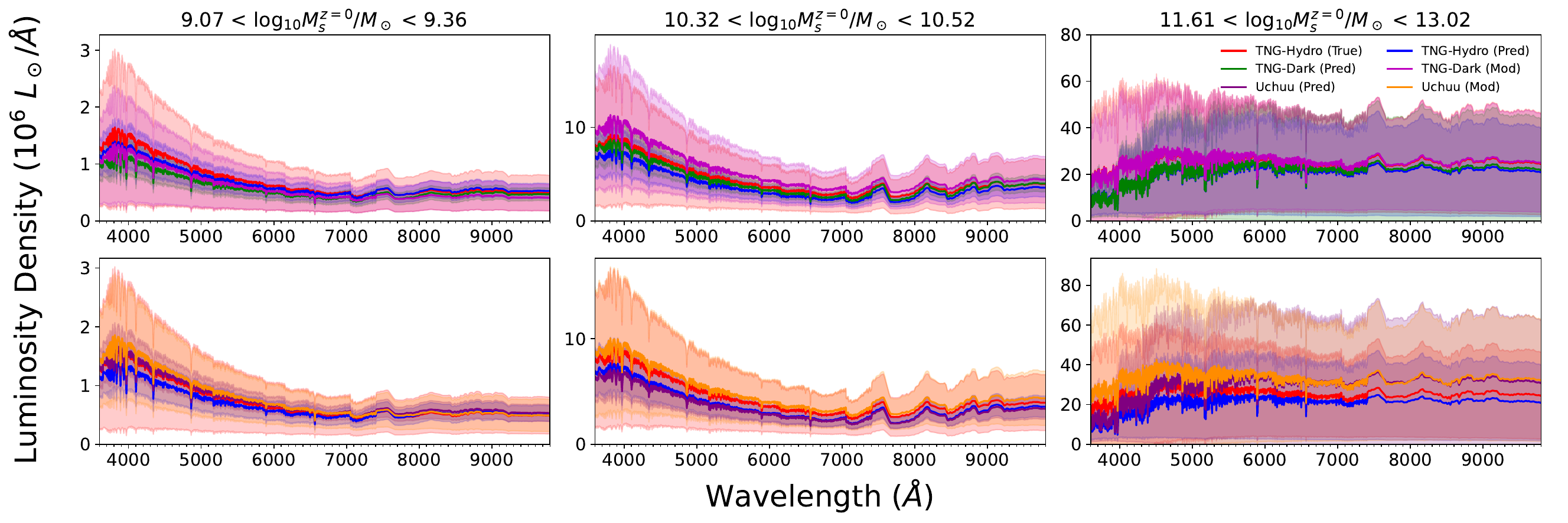}
\vspace*{-20pt}
\caption{For central galaxies, this figure shows the evaluated SEDs in TNG-Dark (upper panels) and Uchuu (lower panels) in contrast with the TNG-Hydro data. For each dark simulation, the spectral luminosities are on average under-predicted slightly, but this is improved by the stochastic modification, which also serves to restore some of the variance in these spectra. Emission lines are not shown in these spectra, for the sake of clearly showing the mean continuum from each simulation.}
\label{fig:Spec}
\end{figure*}

\begin{figure*}
\includegraphics[width=\linewidth]{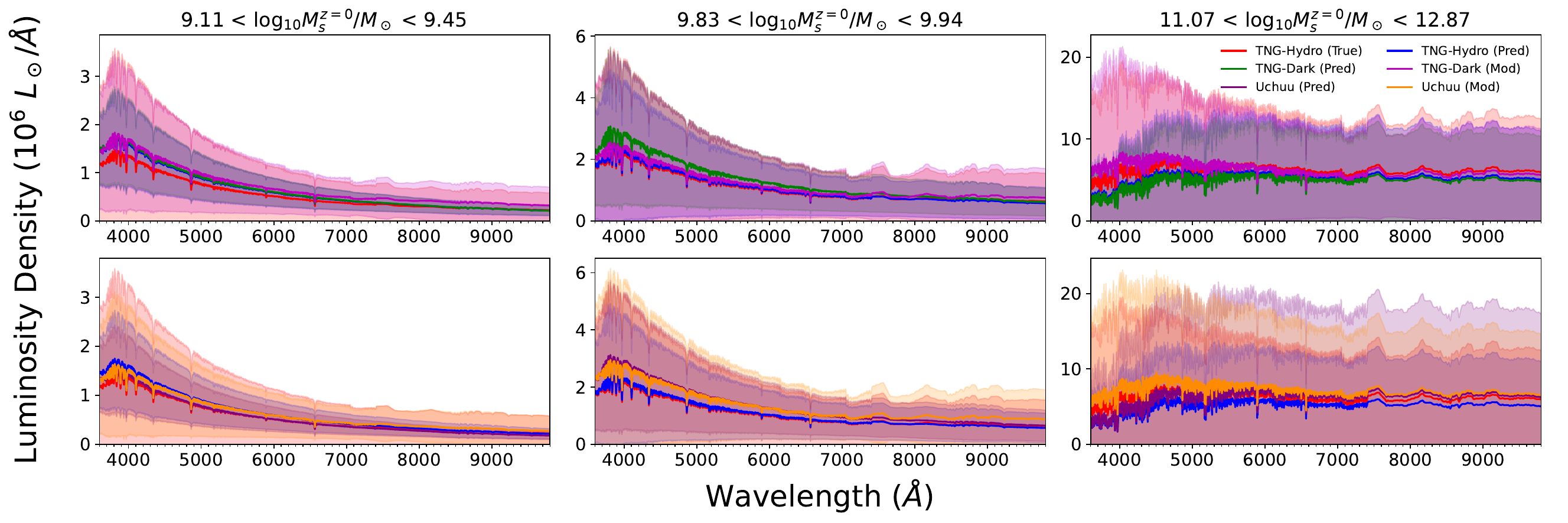}
\vspace*{-20pt}
\caption{For satellite galaxies, this figure shows the evaluated SEDs in TNG-Dark (upper panels) and Uchuu (lower panels) in contrast with the TNG-Hydro data. The two dark simulations are well-matched to the predictions in the hydrodynamic \ac{illtng} simulation, showing similar means and variances in the spectra. This causes the average luminosity to be brighter than needed following modification, but like the central galaxies, the modification recovers the variance originally seen in spectral luminosities. As with \cref{fig:Spec}, emission lines are omitted from the figure for clarity.}
\label{fig:spec}
\end{figure*}

We show in \cref{fig:Spec,fig:spec} the spectral energy distributions of the four simulation datasets in bins of stellar mass, for central and satellite galaxies. In \citetalias{Chittenden} we relate the smaller variance of the predicted SEDs to the lack of variability in star formation histories, and unconstrained implicit features such as merger-driven starbursts and quenching timescales, which are more common to central galaxies.

The mean amplitudes of TNG-Dark spectra are subtly smaller for low and intermediate mass central galaxies, and larger for satellites. The satellite discrepancy can be attributed to higher peaks in star formation histories at such masses. For central galaxies, the cause of this offset is not clear, but given differences in star formation histories and mass-weighted ages, the offset may owe to the underpredicted star formation histories seen in Uchuu predictions. While this appears in both networks, the effect is particularly prominent for satellites, which in addition to the abscence of stellar feedback driven mass loss \citep{Sorini}, may be likened to the lack of environmental harrassment serving to strip the satellite halo following infall \citep{Engler}, as discussed in \cref{sec:sfh}.

The stochastic modification systematically increases these galaxy luminosities as it does with stellar masses, therefore it improves the spectral amplitudes for central galaxies, and not satellites, which additionally exhibit greater variance than the target data when compared with central galaxies. This inconsistency between the spectral modifications of central and satellite galaxies may be explained by the physical processes which act in the effective frequency range of the stochastic modification. While central galaxies tend to quench as a result of mergers and AGN growth, satellite quenching can be driven by the environment imposed by its host galaxy, and can act on short and on long timescales, depending on infall trajectory, satellite-host mass ratio and redshift \citep{Baxter,Mao}.

In Uchuu, the spectra are generally lower in amplitude than TNG-Dark spectra, resulting from weaker star formation histories. One exception is low mass central galaxies, which may be a result of poorly predicted metallicity. In high mass bins, however, the variance in Uchuu spectra is larger than that of any \ac{illtng} data, corresponding to larger variance in mass accretion histories. The large spread of Uchuu overdensities in high mass bins will additionally have misled the network, either by enhancing star formation by associating densities with merger rates, or conversely, suppressing it by association with quenching. In fact, the Spearman coefficient between zero-redshift overdensity and each band magnitude is on average -0.43 for galaxies above $10^{11.5} M_\odot$, compared with -0.09 overall. The stochastic amendment may reshape these large Uchuu galaxy spectra to more closely resemble that of \ac{illtng} galaxies, but of course cannot reduce the variance as needed, only reducing this average Spearman coefficient to -0.38.

\subsection{Galaxy Photometry}
\label{sec:colmass}

\begin{figure*}
\includegraphics[width=\linewidth]{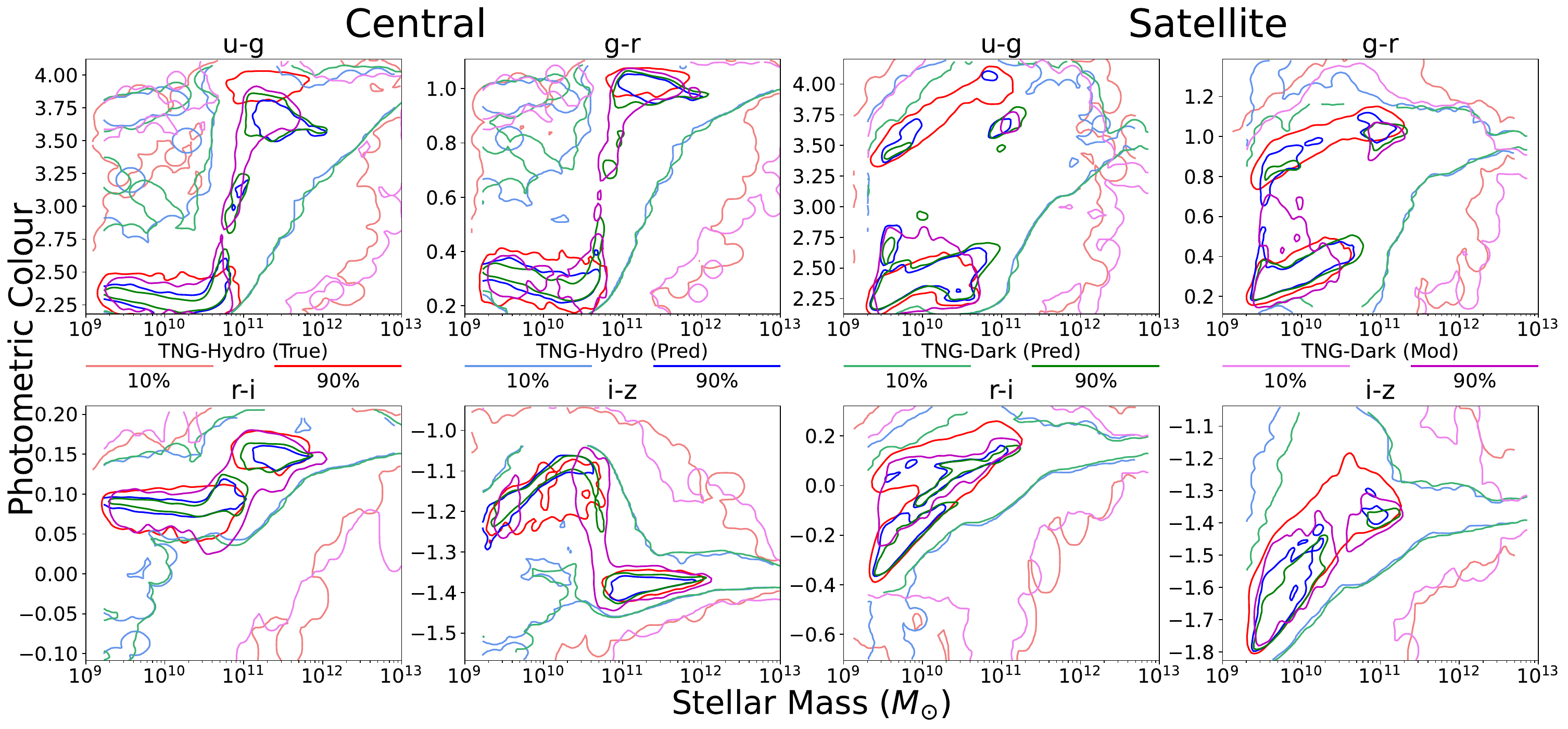}
\includegraphics[width=\linewidth]{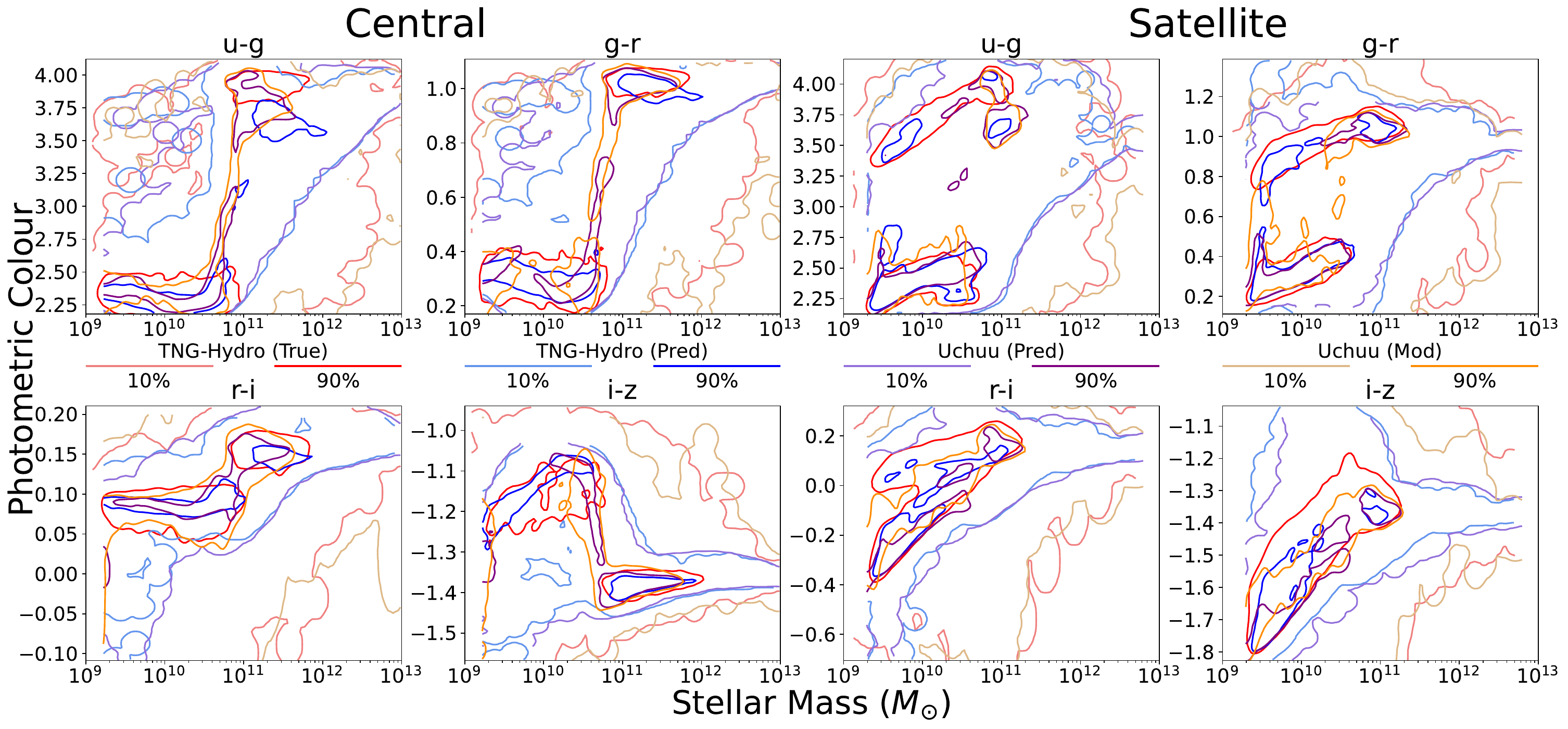}
\vspace*{-20pt}
\caption{Colour-Mass diagrams from each of the four simulation datasets, shown for central galaxies (left panels) and satellite galaxies (right panels), and for four colours taken as the difference between successive SDSS band magnitudes. Contour lines show the $10^\text{th}$ percentiles (light-coloured lines) and the $90^\text{th}$ percentiles (dark-coloured lines) of the 2D histograms of the data. The predicted photometric colour bimodality and its association with mass are shown by all datasets, however the dark simulations show an excess of samples in between the peaks of the colour distributions. The stochastic modification redistributes the colour distributions, recovering the full range of colours seen in the original data, but does not rectify the offset peaks of the distribution, particularly for high frquency bands.}
\label{fig:cmdiag}
\end{figure*}

In \cref{fig:cmdiag}, we show the colour-mass diagrams evaluated from the four simulation datasets, showing for both central and satellite galaxies, the dependence of colours calculated using neighbouring SDSS wavebands on their stellar mass. As in \citetalias{Chittenden}, these results show the bimodal colour distributions of the galaxy populations, and the tendency for high mass galaxies to be redder in colour, to be a feature of all predictions of the neural network. This work showed that the unconstrained emission at UV frequencies has resulted in underpredicted colours for $u$ and $g$ bands, which shows in the dark simulations as well. While the stochastic modification recovers the widths of the bimodal colour peaks, owing to improved variance in star formation and spectral amplitude, it fails to correct the offset peak in $u-g$ colour, suggesting the need to accurately predict recent star formation to accurately constrain key spectral features such as the 4000$\AA$ break. This level of correction is analysed quantitiatively in \cref{sec:KS2}.

In the predictions of the dark simulations, there are some notable differences with respect to the predictions of the hydrodynamical simulation. The ages of high mass galaxies are biased towards high values in both dark simulations, as shown in \cref{fig:mwa}; and correspondingly, a higher fraction of galaxies reside in the "red" peak in these datasets, particularly for $u-g$ and $g-r$ colours. This overabundance of red galaxies is potentially a result of a greater rate of galaxy quenching, which was discussed in \cref{sec:sfh} in relation to the differing morphology of haloes in dark simulations; or the denser environments in dark simulations which serve to quench interacting or infalling galaxies. The stochastic modification partially corrects this by reshaping the star formation histories, making some red galaxies bluer by eliminating excess early star formation, but does not recover quenching tails which would also shift these colours.

Another aspect of the dark predictions is the greater abundance of galaxies in the "green valley": the transition phase between star-forming and quiescent galaxies. This feature applies to both dark simulations, but is particularly prominent in Uchuu. Unlike the colour offset arising from the galaxy age bias, galaxies in this mass range typically have a broader range of calculated magnitudes, which, like the high variance in spectra, may be attributed to variations in the halo mass accretion history and internal dynamics. Furthermore, an effect of the stochastic modification is the enhancement of the green valley population; as the modification is sensitive to a frequency range which captures a specific set of star formation features, the correction it implements may only constitute a partial shift in the photometric colour.

We have shown in \citetalias{Chittenden} that the variability in star formation history, particularly at late times, is an important factor in modelling photometry. Even in cases where the Fourier Transforms of Uchuu galaxies are well predicted, the star formation histories in Uchuu may have lost additional high frequency information from being based on temporal predictors which were interpolated over a sparser time domain in Uchuu, which could not be rectified by a \ac{illtng}-based stochastic modification. Uchuu galaxies in this mass range are also slightly redder on average than TNG-Dark, which corresponds to declining star formation histories in this mass range. In fact, this effect is seen at the approximate mass at which biased galaxy ages begin to appear, suggesting that over-quenching is indeed a prevalent issue. In \cref{sec:KS2}, we see that the improvements of most colour distributions following modification are significant for low mass, preferentially star-forming galaxies, unlike high mass, quiescent galaxies; emphasising the importance of constraining late star formation to accurately predict observable data. However, the modification shows a modest recovery of the colour distribution for massive galaxies in dark simulations, signifying the utility of reshaping their star formation histories.

This modification is nonetheless only able to correct for differences in network predictions which translate to intermediate frequency modes in the star formation history, including supernova feedback but not necessarily slow quenching. Concerning high mass galaxies, \citet{Iyer} show that kinetic AGN feedback introduces a variable star formation component, which in \ac{illtng} contributes significant energy injection only above a threshold black hole mass of $\sim 10^8 M_\odot$; unlike the thermal feedback mode, whose energy yield is tightly correlated with this AGN mass \citep{Zinger}. These distinct modes of AGN feedback may explain why the stochastic modification can only partially amend the geometry of quenched star formation histories; the kinetic feedback resides in the frequency domain of the modification, while the long-timescale thermal feedback is closely related to AGN mass. While AGN mass is not predicted directly by this model, it is closely related to star formation history, and so the effects of thermal AGN feedback will be influenced by the established shortcomings of the fiducial neural network predictions, in both the hydrodynamic and the dark simulations.

\section{Discussion}
\label{sec:disc}

The machine learning model used to predict the star formation and stellar metallicity histories of \ac{illtng} galaxies has performed modestly at reproducing the same results in pure dark matter simulations, having obtained similar quantitative relations between galaxies and haloes, and relating physical galaxy properties to observational quantities as in the hydrodynamical \ac{illtng} simulation. Despite this, a number of discrepancies exist between the different predictions, resulting from differences in the growth and interactions of haloes in N-body simulations compared with the full physics, differences in the calculation of halo properties in the simulation data, and the degree of improvements made by stochastic modification. In this discussion, we identify the important differences between these simulations and how the model may be modified in future to suit high volume N-body simulations while maintaining an accurate characterisation of the galaxy-halo connection.

\subsection{Resolution Effects}
\label{sec:reseffects}

We have shown that the lower resolution of the Uchuu simulation compared with the \ac{illtng} simulations has worsened the quality of predictions of low mass and slowly accreting galaxies, with considerable errors in the masses and metallicities of low mass galaxies; particularly satellite galaxies, whose sampling was already compromised by our quality cuts. One such effect is that the truncated power spectrum suppresses the coalescence of simulation particles into low mass haloes, such that low mass haloes take longer to germinate and grow. Low mass samples in Uchuu are evidently unreliable for causal galaxy-halo modelling. The \ac{sfh} and \ac{zh} Fourier Transforms predicted by this model are poorly predicted for these haloes as well, indicating that resolution differences additionally affect our stochastic methodology.

The limits of computational resources required to generate a high-volume, high-resolution simulation has been a long-standing issue in this field. \citet{Li} demonstrate a machine learning model which can enhance the matter power spectrum in N-body simulations and generate accurate snapshot data from low-resolution simulation images, which accurately replicate halo substructures and correlation functions below the limit of said low-resolution simulations \citep{Ni}, while tracing realistic merger histories across simulated high-resolution snapshots \citep{Zhang}. Given the importance of the properties of progenitor subhaloes, such as the mass and metallicity of incoming gas and stars, these high resolution merger trees may be of practical use when enhancing the Uchuu data in this work; however, as the matter power spectra, distributions and velocity fields can be accurately enhanced this way, these properties may be inferred from the predicted environments.

Our results show that haloes in Uchuu of similar mass to the lowest mass haloes in \ac{illtng} are substantially affected by the difference in resolution, and thus, it may be worthwhile in a future study to apply the neural network to Shin-Uchuu: a smaller but higher resolution run of the Uchuu model, to restore low mass galaxies; or to enhance the Uchuu snapshots and merger trees based on the superior Shin-Uchuu resolution and the methods of \citet{Li}. As the growth of large scale structure will influence the distribution of galaxies and the properties of their environments, the latter may be a more suitable method for creating self-consistent mock catalogues on both gigaparsec and sub-kiloparsec scales.

\subsection{Alternative Models}
\label{sec:UchuuUM}

\subsubsection{Uchuu-UniverseMachine}

\citet{UchuuUM} use the \ac{UM} model \citep{Behroozi} to compute galaxy formation histories in the Uchuu simulation, which is used to recover statistics such as stellar mass functions and number density profiles which show reasonable agreement with observational stellar mass and luminosity functions \citep{Behroozi2019}. \ac{UM} is an empirical model, relying on MCMC optimisation of star formation rates using prior relations of star formation rates and quenched fractions to the halo rotation curve. The model was able to qualitatively reproduce the environmental dependence of star formation without explicit implementation of the environment, yet this implies that the star formation in dense environments is only supported by halo mass accretion. \ac{UM} does not model metallicity histories, which we have shown in \citetalias{Chittenden} to be more dependent on environmental quantities, and thus our model may be more suitable for modelling the dependence of chemical enrichment in high fidelity mocks. On the contrary, as quenched fractions are a direct parameter of the \ac{UM} model, the colour bimodality in these mocks is likely to be more accurate, and so \ac{UM} may be more suitable for observational statistics.

A predictive, self-consistent model such as ours may prove complementary to empirical models such as \ac{UM} when it comes to producing high-fidelity mocks, as it can be used to causally model the growth of galaxies over time based upon the halo and environmental quantities driving the physics of the galaxy-halo connection, and may be modified to predict galaxy properties, for instance, spiral-bar structures. Yet, our machine learning model and stochastic modification have shown to be sensitive to effects which come from translation to a pure dark matter simulation, such as the biased growth of internal dynamics of the halo, resolution effects such as the delayed collapse and virialisation of haloes, and potentially, nuances in the calculation of key variables such as halo mass and radius due to the use of different halo finder algorithms; which may, for instance, identify different structures in the halo centre, or different boundaries enclosing the total halo mass \citep{Onions}. Thus, it may prove that an N-body simulation with resolution enhancements and consistent halo definitions with the training data will be necessary for accurate self-consistent mocks.

\subsubsection{Gaussian Process Stochasticity}

Our stochastic modification to the star formation and metallicity histories has proven valuable in recovering the short-timescale features which were absent from the fiducial neural network model, corresponding to relevant phenomena such as starbursts and baryon cycling. It is nonetheless difficult to disentangle the physical drivers of these stochastic features, which have different degrees of influence depending on a galaxy's mass, age and environment. To analyse the relative contribution of these dynamical processes to observable features such as the H-$\alpha$ flux and $D_n$4000 index, one might incorporate a Gaussian process formulism as seen in \citet{Iyer2}, where the stochastic modes of star formation are parameterised by three analytic power spectra with controlled effective timescales.

Our stochastic phase selection method presented in \citetalias{Behera} considers the cross-correlation of modes of halo and galaxy growth, and so the evolution of the host halo may provide some insight into constraining these timescales for individual galaxies, if these two stochastic analyses can be reconciled. This may require more historical information than what was presented in our companion paper, considering that the gas inflow and cycling studied by \citet{Iyer2} are potentially sensitive to environments; but if the stochastic parameters can be infefrred solely from properties of the halo and environment, they can contribute to self-consistent predictions in N-body simulations.

\subsection{Alternative Variables}
\label{sec:altvars}

The variables used in the design of our neural network were meant to characterise the growth, structure and environment of haloes in a manner which was as immune to the effects of baryons and different resolutions and halo finders as possible. However, the results have shown that these differences have nonetheless produced small to substantial differences in the statistics of galaxies predicted using pure dark matter simulations with respect to the hydrodynamical \ac{illtng} model. Thus, the inclusion of as-yet unused variables in a future rendition of the neural network model may constitute more suitable measures of the galaxy-halo connection for use in a pure dark matter model.

As mentioned in \cref{sec:UchuuUM}, the use of a consistent halo finder in the hydrodynamical and N-body simulations under consideration may be necessary to avoid biasing predictions due to the different identification of halo mass and substructure. \citet{Onions} state that while many halo finder algorithms are alike in their capability of identifying halo structures, Rockstar is the only algorithm in this study which accurately resolves substructures in highly dense regions such as the halo centre, which can influence the structural variables $R_\frac{1}{2}$ and $\tilde{v}_\text{vir}$; shown in \citetalias{Chittenden} to influence the \ac{shmr}s of both central and satellite galaxies. The planned \ac{illtng} halo catalogue based on the Rockstar halo finder may be used to train an "Uchuu-friendly" neural network model in future research, effectively eliminating this potentially appreciable biasing factor.

Another quantity which differs between the simulations in this study is overdensity, which in Uchuu is computed using halo coordinates and masses in lieu of similar subhalo information as in \ac{illtng}. Overdensities may be smoothed according to a Gaussian kernel to produce a continuous, position-dependent density field as in \citet{Chen}, which may be tuned for each simulation according to their resolution or their density tracers. This method was used similarly to create a position-dependent tidal field tensor, which the authors show to correlate strongly with halo assembly bias.

For satellite galaxies, the environment they experience in the vicinity of their host halo dominates their behaviour in the satellite phase, and therefore the properties of the host are additionally important. In \citetalias{Chittenden} we mention how the explicit inclusion of further satellite host properties may have been beneficial to the model; examples include formation time \citep{Artale}, distance from the halo centre \citep{Engler, Montero-Dorta2} and angular momentum \citep{Bose}. The location of the satellite with respect to the host, or of a central halo with respect to a local cluster, would constitute a tracer of the local environment, and thus could be a valuable measure of effects such as gas stripping and tidal disruption.

As the skew parameter, being mass-independent and calculated using halo positions, does not differ substantially between any of the simulations used in this work, a purely position or velocity dependent environmental metric may be a feasible solution to the bias introduced by basing environments on overdensities. However, as the skew is a measure of halo-halo interactions over time, something not characterised by smooth fields, it remains a valuable parameter of this model. Yet as mentioned in \cref{sec:zh}, an effectively random error may be introduced due to the lack of low-mass objects in a low resolution simulation, which would be particularly detrimental to the major interactions inferred from high or low skews.

As for the stochastic modification, while we use stellar mass to bin samples and recover separate phase distributions, we have tested the modification with an additional halo mass binning, though this introduced errors in cases of low sample size, such as with the most massive haloes. It should be stressed here that the stochastic modification does not perfectly reconstruct important summary statistics such as the \ac{mzr}, and sampling phases with respect to environment may improve this result, provided that these are well sampled. Otherwise, combining results from simulations on different scales, such as TNG50 \citep{Nelson2019,Pillepich2019} for small scales and MillenniumTNG \citep{BoseMTNG,HernandezAguayo,Pakmor} for large scales, may provide enough data to justify multidimensional phase selection.

\section{Conclusions}
\label{sec:conc}

In this work, we have compared the quality of predictions of galaxy star formation and metallicity histories in pure dark matter simulations, based on the semi-recurrent neural network described in \citet{Chittenden}, and the stochastic star formation modelling of \citet*{Behera}. We have compared our original predictions with cross-matched haloes from the dark equivalents of the \ac{illtng} simulations on which the model was trained, evaluating the effects on predictions due to the abscence of baryonic processes; and we have applied the model to similar haloes from the Uchuu N-body simulation, to evaluate the effects of alternative halo definitions, and the lower mass resolution of the simulation.

Our findings can be summarised as follows:

\begin{itemize}
\item Important input properties such as the mass accretion history of a halo are similar between the simulations under consideration, for haloes of most masses and mass accretion gradients, as defined in \citet{Montero-Dorta} and \citet{Shi}. Nevertheless, we show in \cref{sec:mhdot} that the mass accretion histories of TNG-Dark haloes are exaggerated, potentially as a result of the lack of stellar feedback shaping the haloes. This has noticable, similar influence on the star formation histories of low mass galaxies, as shown in \cref{sec:sfh}. As this difference in mass accretion histories arises at early times, the consequential effect on the star formation histories applies for most of the time domain of the simulation, due to the recurrent design of the neural network.
\item The network quantities relating to internal structure and dynamics: the half-mass radius and virial velocity of the halo, are similarly affected by the lack of baryons, but are more profoundly affected by the lower resolution of the Uchuu simulation, which becomes apparent in \cref{sec:rhalf}. This causes the germination and initial growth of haloes to be delayed and for the mass accretion and concentration of Uchuu haloes to appear smaller than their \ac{illtng} equivalents. For low mass and slowly growing haloes, these effects are dramatic, due to the sensitivity of these results to the simulation resolution.
\item We show largely similar neural network predictions to the original, hydrodynamical predictions in \cref{sec:pred}, which recover similar shape and scatter to the hydrodynamical \ac{shmr} and \ac{mzr}. Despite this success, the severely underpredicted growth of low mass haloes in Uchuu leads to poor predictions of the star formation and metallicity histories of low mass galaxies. Conversely, TNG-Dark results are overpredicted due to excess mass accretion at early times, which leads to an effective overabundance of star-forming gas. In each case, the lowest mass galaxies in any dark matter simulation are the least accurate predictions.
\item With the added stochastic modification, we make significant improvements to key statistics of the dark simulations, recovering the scatter of the mass-metallicity relation and correcting the spectral luminosity and photometric colour distributions of predicted galaxies. While this amendment is shown to fruitfully reproduce the missing features in the hydrodynamic \ac{illtng} simulations in our companion paper, it also rectifies systematic distortions in the geometry of the star formation and metallicity histories in these pure dark matter simulations, providing more accurate predictions for future N-body mocks. However, it does not amend the issues which largely apply to low mass, under-resolved objects, where the predicted Fourier Transforms are under-predicted, thereby failing to provide adequate information to compute realistic baryonic data. The modified data is therefore just as sensitive to resolution effects as the fiducial predictions.
\item In both dark simulations, the number of quenched galaxies predicted by the neural network is larger than the hydrodynamical result. In TNG-Dark, this is a result of the difference in structural parameters, which may be attributed to AGN feedback. In Uchuu, this is an effect of higher overdensities, which owe to the use of halo tracers rather than subhaloes. This excess quenching corresponds to a greater abundance of photometrically red galaxies, which we show in \cref{sec:obs}. In Uchuu, the abundance of red galaxies is further attributed to underpredicted Fourier amplitudes, and possibly interpolation over a coarser time domain, resulting in greater information loss regarding time variations in their star formation history. The stochastic modification serves to redistribute the photometric colours by reshaping the star formation histories, but does not amend differences in the shapes of quenching tails, leading to some "partly corrected" photometry, and an overabundance of green valley galaxies. The spectroscopic and photometric statistics of the dark matter simulations nonetheless show similar characteristics to the original \ac{illtng} results, yet this could be improved by explicit constraints on star formation at the lowest redshifts.
\end{itemize}

This paper has shown that the neural network model we have developed is a highly practical tool for emulating galaxies into N-body simulations using the learned physics of the galaxy-halo connection. However, the shortcomings of the study illustrate the necessity for an N-body simulation of similar mass resolution and halo properties to compute accurate mock surveys on all mass scales. This may be achieved in future work by redesigning the predictive model to measure halo structure and cosmic environment using identical halo finders, metrics which are independent of the simulation, or by enhancement of existing low-resolution haloes and merger trees using machine learning or similar methods.

By obtaining results in the Uchuu simulation with similar characteristics to the \ac{illtng} simulation suite on which the model was trained, we have shown that it is possible to produce a physical, self-consistent model of the galaxy-halo connection on present-day, gigaparsec-scale N-body simulations. This may be a pathway to a future of AI-based simulations which entail a physical explanation of galaxy evolution on cosmological and substructure levels simultaneously, and corresponding observational data to complement the most ambitious galaxy surveys to date, for a fraction of the computational cost of a hydrodynamical or semi-analytic model of the same level of detail.

\section*{Acknowledgements}

The UKRI Science and Technology Facilities Council supported this study under grant ID ST/T506448/1, which the authors gratefully acknowledge. We appreciate the IllustrisTNG project allowing us access to their data and JupyterLab environment, and the Uchuu and YTree projects for access to Uchuu halo merger trees. We also thank Katarina Kraljic for assistance with modelling the cosmic web in Uchuu, and we thank Daniele Sorini and Kartheik Iyer for useful discussion on the role of gas-rich progenitors on developing halo shapes, and the influence of different modes of baryon cycling and AGN feedback in galaxy evolution models, respectively. Finally, we thank the anonymous reviewer for constructive feedback which helped to improve the quality of this manuscript.

\section*{Data Availability}

The data relating to this publication and associated works are available on an online \href{https://doi.org/10.5281/zenodo.15589166}{Zenodo} repository, while the code is available on an online \href{https://github.com/hgc4/TNG-Networks/}{GitHub} repository.

\bibliographystyle{mnras}
\bibliography{bib}

\newpage

\appendix
\crefalias{section}{appendix}

\section{Quantitative Validation of Modified Predictions}
\label{sec:quant-analysis}

\subsection{Comparing Binned Statistics}
\label{sec:KS}

A key motivation for introducing stochastic corrections into our model is the limited capacity of our deterministic neural network model to reproduce the full diversity of galaxy evolution, such as in regimes where stochastic processes such as environmental influence or feedback-driven variability play a dominant role. This appendix presents a quantitative validation of the improvements introduced by our stochastic modification.

We assess model performance by comparing predicted quantities from each of the three simulations studied in this work, both with and without the stochastic term, against their counterparts in the original TNG data. To do this, we employ three main statistical tools: binned means, standard deviations, and Kolmogorov-Smirnov (KS) statistics. These metrics offer complementary insights: the mean traces the average behaviour of a given galaxy property (e.g. stellar mass or metallicity), while the standard deviation captures the population-wide scatter, offering a proxy for the intrinsic diversity of galaxy formation histories. Together, they allow us to assess how well the model recovers both central trends and variance across the galaxy population.

To go beyond summary statistics, KS testing provides a non-parametric measure of similarity between two cumulative distributions. For each halo or subhalo mass bin, we compute the KS statistic: the maximum vertical distance between two cumulative distribution functions, between the predicted and original (TNG) distributions of each quantity; where lower KS values correspond to better agreement with the target distribution. We further define two metrics to quantify the effect of the stochastic correction, the KS statistic reduction:

\begin{equation}
\Delta_\text{KS} = \text{KS}_\text{NN} - \text{KS}_\text{NN+Mod},
\end{equation}and the percentage KS reduction:

\begin{equation}
P_\text{KS} = 100 \times \frac{\Delta_\text{KS}}{\text{KS}_\text{NN}} = 100 \times \frac{\text{KS}_\text{NN} - \text{KS}_\text{NN+Mod}}{\text{KS}_\text{NN}}.
\end{equation}

With these metrics, a large positive value of $\Delta_\text{KS}$ or $P_\text{KS}$ indicates a significant improvement due to the stochastic modification, while small or negative values suggest little to no benefit.

While this metric works well for stellar mass and stellar metallicity, the same cannot be said for photometric colours, due to shifts in their bimodal peaks, which compromise the value of a difference in two functions at a single point. Instead, we define the root mean square (RMS) statistic as the root mean square of the integrated vertical difference between two cumulative distributions. For a photometric colour $x$, the RMS statistic for the fiducial model, $\text{RMS}_\text{NN}$, is defined:

\begin{equation}
\text{RMS}_\text{NN} = \sqrt{\int_{x_\text{min}}^{x_\text{max}} \bigg( \text{CDF}_\text{NN} (x) - \text{CDF}_\text{TNG} (x) \bigg)^2 dx},
\end{equation}and $\text{RMS}_\text{NN+Mod}$ is defined equivalently for the modified data, while $\Delta_\text{RMS}$ and $P_\text{RMS}$ are defined in the same manner as their KS statistic counterparts.

For stellar mass and metallicity, we analyse the mean and standard deviation across halo/subhalo mass bins, in addition to KS statistics. For photometric colours, we consider only RMS statistics, as their bimodal distributions make the mean and standard deviation less informative.

Statistical comparisons are carried out separately for central and satellite galaxies, using the following four bins of halo and subhalo mass:

\begin{itemize}
\item High Halo/Subhalo Mass: \hfill $\log_{10} M_h / M_\odot > 13$
\item Intermediate Halo/Subhalo Mass: \hfill $12 < \log_{10} M_h / M_\odot < 13$
\item Low Halo/Subhalo Mass: \hfill $\log_{10} M_h / M_\odot < 12$
\item Very Low Halo/Subhalo Mass: \hfill $\log_{10} M_h / M_\odot < 11.5$
\end{itemize}

This framework allows us to isolate trends across different mass regimes and galaxy types, providing a robust test of the stochastic correction's effectiveness in restoring physically motivated diversity to galaxy property predictions.

\subsection{Galaxy Properties}
\label{sec:KS1}

\begin{figure*}
\includegraphics[width=\linewidth]{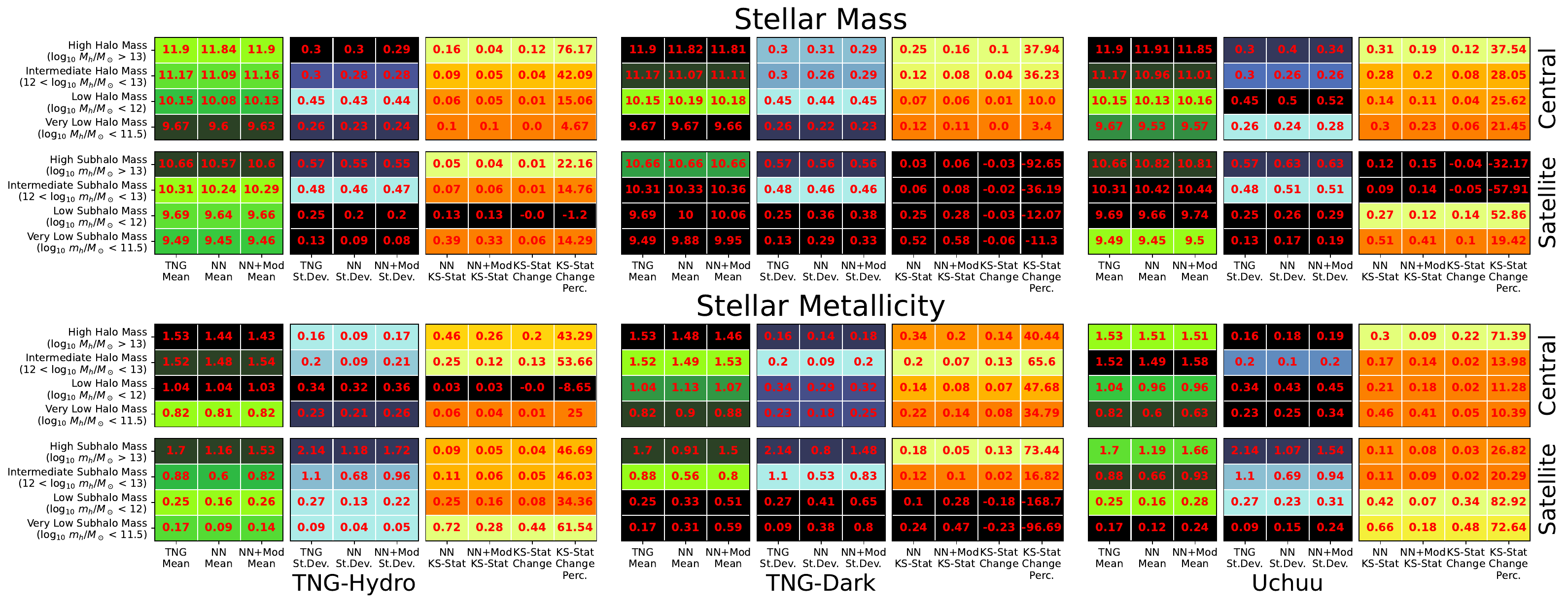}
\vspace*{-20pt}
\caption{KS statistic analysis of predicted stellar mass (top) and stellar metallicity (bottom) across TNG-Hydro (left column), TNG-Dark (centre), and Uchuu (right) simulations. Each grid shows results across bins of halo mass and subhalo mass, separately for central and satellite galaxies, whereas the colour scale of each row indicates better agreement with the TNG-Hydro data, and is scaled according to either the difference of the modified mean from its target, equivalently for standard deviation, or is simply scaled by the percentage KS reduction. Rows in the grid with no improvement from the modification are strictly coloured black. Columns in the first block of each grid, with a green colour scale, show the TNG reference mean, the fiducial neural network (NN) prediction and the modified prediction (NN+Mod). The second block, whose colour scale is blue, shows the same for standard deviation, while the third, coloured gold, shows KS statistics comparing each prediction to the original TNG values. The final two columns in the third block indicate the absolute and percentage change in the KS statistic after modification. The stochastic modification generally leads to improved agreement, particularly in stellar metallicities of satellite galaxies. Stellar mass values in this figure are shown as logarithmic quantities, and both stellar masses and metallicities are in solar units.}
\label{fig:KS}
\end{figure*}

\Cref{fig:KS} shows how stochastic modification improves predictions of stellar mass and metallicity, using KS statistics across bins of halo and subhalo mass, for both centrals and satellites in TNG-Hydro, TNG-Dark, and Uchuu.

Stellar mass predictions are already accurate in most regimes, so improvements are modest. The most notable gains occur in high and intermediate-mass central galaxies, with up to a 76\% reduction in KS error for TNG-Hydro; reflecting the correction of missing AGN-driven variability. Low-mass galaxies, particularly in Uchuu, show higher KS values due to poor resolution; percentage improvements here are less meaningful. For example, the KS value of 0.51 for low-mass Uchuu satellites highlights a severe mismatch, though this owes to amplitude suppression by poor resolution, as opposed to distorted SFH geometries in TNG-Dark.

Metallicity predictions benefit much more broadly. The stochastic model restores scatter lost in the deterministic predictions, particularly in the mass-metallicity relation. This leads to 40-60\% KS reductions in TNG-Hydro and TNG-Dark, and up to 71\% for high-mass Uchuu centrals, driven by improved modelling of early star formation and enrichment events. The exception is low-mass Uchuu satellites, where resolution limits compromise the accuracy of the \ac{mzr}, regardless of modification.

In summary, the stochastic correction offers targeted improvements: strong gains in metallicity across all regimes, and notable stellar mass improvements in high-mass systems, especially where AGN feedback plays a dominant role.

\subsection{Observable Quantities}
\label{sec:KS2}

\begin{figure*}
\includegraphics[width=\linewidth]{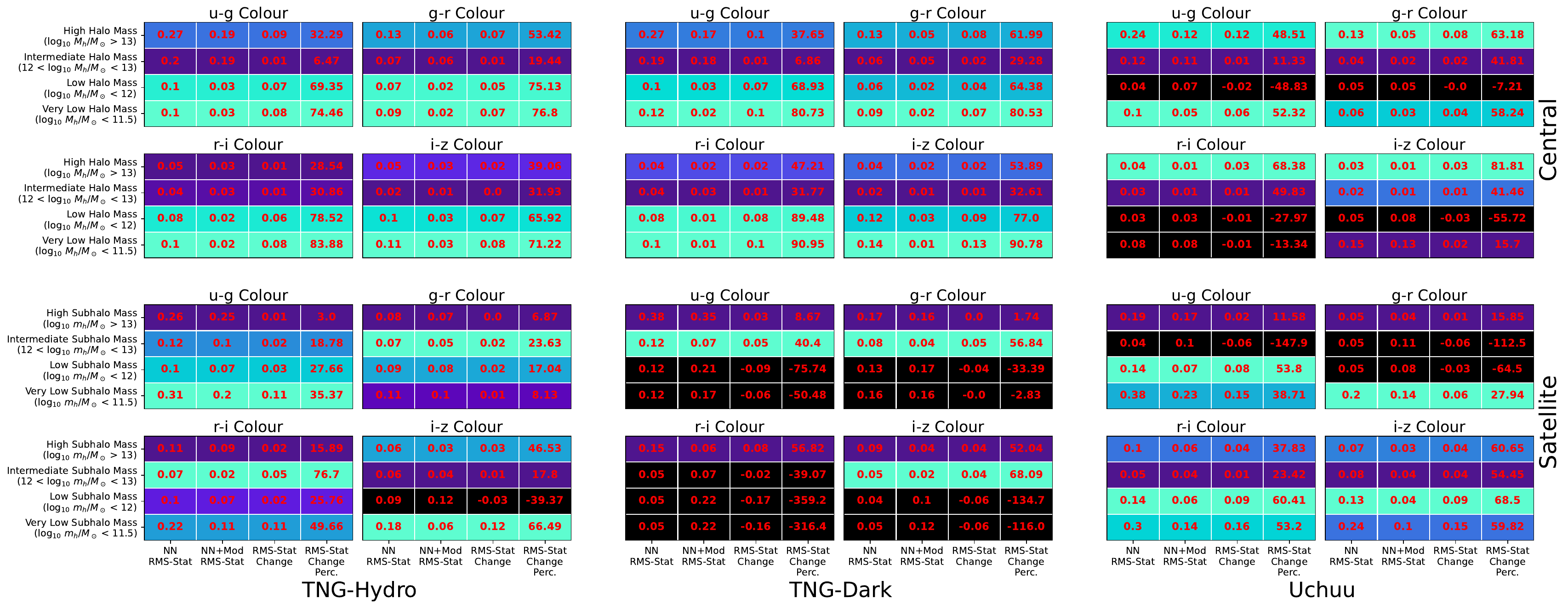}
\vspace*{-20pt}
\caption{RMS statistics and stochastic RMS reductions for rest-frame photometric colours, for central galaxies (top) and satellites (bottom). RMS statistics are calculated by comparing predicted colour distributions from the fiducial and modified neural network outputs to those in the TNG simulation. As with intrinsic properties, the stochastic modification improves the agreement with TNG in most bins, with the most pronounced improvements observed in satellite populations, where environmental diversity is less well captured by deterministic models. The brighter row colours in this grid also indicate better agreement with the TNG-Hydro reference data.}
\label{fig:RMS}
\end{figure*}

In addition to intrinsic properties, we assess the impact of the stochastic correction on observable quantities, focusing on the four photometric colours previously consdered in this work. As shown in \cref{fig:RMS}, RMS statistics indicate consistent improvements across halo and subhalo mass bins, provided that the predicted Fourier transforms are accurate.

The largest RMS reductions occur in low-mass galaxies, whose colours are highly sensitive to recent star formation. While the fiducial model struggles to capture this variability, the stochastic modification reintroduces variance, restoring diversity from bursts and quenching; and yields colour distributions which are more consistent with the hydrodynamic reference. However, in intermediate to high-mass systems, while the width of the colour distribution improves, the peak offset remains. This suggests that recent star formation episodes are still not explicitly captured by the correction, highlighting a limitation of the approach.

Percentage RMS improvements are often marginally greater in the dark matter simulations. At low mass, this reflects correction of the red fraction, countering the early-biased SFHs in TNG-Dark. At higher masses, moderate improvements suggest partial AGN-driven variability is recovered, particularly in Uchuu. However, the correction still fails to reproduce long-timescale features like quenching tails, limiting improvements in massive systems.

Overall, these results confirm that stochasticity improves photometric predictions, especially in regimes shaped by short-timescale variability; enhancing the realism of simulated galaxy populations.

% Don't change these lines
\bsp	% typesetting comment
\label{lastpage}
\end{document}